\documentclass[aps,reprint,nofootinbib,longbibliography,onecolumn]{revtex4-2}
\usepackage{graphicx,multirow}
\usepackage{subcaption}
\usepackage{float}
\usepackage{braket}
\usepackage{amsmath}
\usepackage{amssymb}
\usepackage{amscd}
\usepackage{latexsym}
\usepackage{slashed}
\usepackage{color, xcolor}
\usepackage{graphicx}
\usepackage{ulem}
\definecolor{orange}{cmyk}{0,0.5,1,0}
\usepackage{verbatim}
\usepackage{rotating}
\usepackage{cancel}
\usepackage{caption}
\usepackage{hyperref}
\usepackage{lineno}
\usepackage{makecell}
\usepackage{natbib}

\hypersetup{colorlinks=true, linkcolor=blue, citecolor=red,filecolor=magenta}

\begin{document}

\title{Leptonic CP-violation in the sneutrino sector of the BLSSM with Inverse Seesaw}

\author{Arindam Basu} 
\email{arindam_basu@srmap.edu.in}
\affiliation{Department of Physics, SRM University AP, Amaravati 522240, Andhra Pradesh, India}

\author{Amit Chakraborty} 
\email{amit.c@srmap.edu.in}
\affiliation{Department of Physics, SRM University AP, Amaravati 522240, Andhra Pradesh, India}

\author{Yi Liu} 
\email{Yi.Liu@soton.ac.uk}
\affiliation{School of Physics and Astronomy, University of Southampton, Southampton SO17 1BJ, UK}

\author{Stefano Moretti} 
\email{stefano@soton.ac.uk; stefano.moretti@physics.uu.se}
\affiliation{School of Physics and Astronomy, University of Southampton, Southampton SO17 1BJ, UK}
\affiliation{Department of Physics and Astronomy, Uppsala University, Box 516, SE-751 20 Uppsala, Sweden}

\author{Harri Waltari}
\email{harri.waltari@physics.uu.se}
\affiliation{Department of Physics and Astronomy, Uppsala University, Box 516, SE-751 20 Uppsala, Sweden}

%=============================

\begin{comment}
    
\begin{center}

{\large \bf Leptonic CP-violation in the sneutrino sector of the BLSSM with Inverse Seesaw} \\    
\vskip 0.6cm
Arindam Basu,$^{a}$%\footnote{shreechetac@srmap.edu.in}
, 
Amit Chakraborty$^{a}$%\footnote{aamit.phy@gmail.com} 
,  
Yi Liu,$^{b}$%\footnote{saunak100@gmail.com}
,  
\vskip 0.2cm
Stefano Moretti,$^{b,c}$%\footnote{saunak100@gmail.com}
,  and 
Harri Waltari,$^{c}$%\footnote{saunak100@gmail.com}
\vskip 0.6cm

{$^a$  Department of Physics, SRM University-AP, Amaravati, Mangalagiri 522240, India}
%
\vskip 0.1cm
{$^b$ 
School of Physics and Astronomy, University of Southampton, Southampton SO17 1BJ, UK
} \\
\vskip 0.1cm
{$^c$ 
Department of Physics and Astronomy, Uppsala University, Box 516, SE-751 20 Uppsala, Sweden
} \\
\end{center}
\end{comment}

%============================

\begin{abstract} 
{We study CP violation (CPV) in the sneutrino sector within the B-L extension of the Minimal Supersymmetric Standard Model (BLSSM), wherein an inverse seesaw mechanism has been implemented. CPV arises from the new superpotential couplings in the (s)neutrino sector, which can be complex and the mixing of CP-eigenstates induced by those couplings. CPV leads to asymmetries in so-called T-odd observables, but we argue that such asymmetries also lead to a wider distribution of those observables. We look at a final state where a sneutrino decays to a lepton, two jets, and missing transverse momentum at the Future Circular Collider operating in hadron-hadron mode at $100$ TeV and with a luminosity of 3~ab$^{-1}$. In order to exclude the CP conserving scenario we need to improve traditional analysis by introducing boosted decision trees using both standard kinematic variables and T-odd observables and we need $Z^{\prime}$ boson not too much above current bounds as a portal to produce sneutrinos efficiently.}
\end{abstract}

\maketitle

\section{Introduction}

Non-zero neutrino masses have been observed by several experiments, in fact, such observations represent one of the strongest hints for physics Beyond the Standard Model (BSM), as in the latter scenario neutrinos are assumed to be massless, for two reasons. Firstly, there are no Right-Handed (RH) neutrinos as Dirac fermions to which weak currents can couple. Secondly, the SM strictly conserves the Baryon minus Lepton ($B-L$) number. To address this, though, the SM can be extended by a new $B-L$ symmetry, e.g., with a $U(1)$ gauge structure, allowing for a Majorana nature of neutrinos. The emerging $U(1)_{B-L}$ extended SM, or BLSM (see Ref.~\cite{Khalil:2012gs,Khalil:2013in} for a review)
thus leads to new testable signals, including heavy neutrinos, a neutral gauge boson $Z^\prime$ and a new Higgs state that breaks the $U(1)_{B-L}$ gauge group \cite{Wetterich:1981bx,Buchmuller:1991ce,Khalil:2007dr,Basso:2008iv,Basso:2009gg,Basso:2009zz,Basso:2009hf,FileviezPerez:2009hdc,Basso:2010dq,Basso:2010si,Basso:2010jm,Basso:2010tv,Basso:2010jt,Li:2010rb,Abdallah:2015uba,Abdallah:2015hma,Accomando:2016rpc,Accomando:2017qcs,Gutierrez-Rodriguez:2024nny,Martinez-Martinez:2023qjt,Iso:2009nw,Dev:2021qjj}. 

A seesaw mechanism provides an elegant explanation for neutrino masses. In a canonical seesaw mechanism, such as, e.g., Type-I seesaw, RH neutrinos acquire mass at the $B-L$ symmetry breaking scale (TeV or so). Consequently, the corresponding Yukawa couplings are constrained to be  smaller than $\mathcal{O}(10^{-6})$ \cite{Khalil:2007dr}. The inverse seesaw mechanism \cite{Khalil:2006yi,Khalil:2010iu} can allow for larger Yukawa couplings, making this scenario more feasible to test at the Large Hadron Collider (LHC). Specifically, by introducing two SM gauge singlet fermions, the light neutrino obtains its mass when one of the two singlets interacts with the RH neutrino. Furthermore, the other singlet can serve as a Dark Matter (DM) particle, for which there is no viable candidate in the SM \cite{El-Zant:2013nta} (see also Ref. \cite{Basso:2012gz}).

There are some hints of a non-zero CP phase in the neutrino sector \cite{T2K:2019bcf,NOvA:2021nfi}, although the results come with large error bars. If Charge and Parity Violation (CPV) arises from the dynamics that generates neutrino masses, it may eventually manifest itself in specific kinematical observables, which appears as a difference in the angular distributions of multi-body decay processes of a particle and corresponding antiparticle. Triple-Product Asymmetries (TPAs) are proportional to the scalar triple product of the generic form: 
\begin{equation}\label{eq:TPA}
    {\rm TPA} = \Vec{v}_1 \cdot (\Vec{v_2} \times \Vec{v_3}),
\end{equation}
where the vectors $\Vec{v}_i$ are the three-momenta of final state particles. TPAs are generally T-odd observables, and those that can violate CP symmetry are known as `true' TPAs. Therefore, TPAs can serve as an indicator of CPV BSM physics with a relatively clean background from the SM (only onset by Cabibbo-Kobayashi-Maskawa (CKM) phases), which can lead to a deeper understanding of the CP properties of such new interactions \cite{Gronau:2011cf,Bevan:2014nva}. In this work, we show that besides the non-zero expectation value of the triple product, also the width of its distribution may serve as an indicator of CPV.

Another flaw of the SM is the so-called hierarchy problem, which can, however, be elegantly solved by Supersymmetry (SUSY) \cite{Moretti:2019ulc}, so it is tempting to combine the $U(1)_{B-L}$ symmetry with inverse seesaw and SUSY. This leads to the $B-L$ Supersymmetric SM with IS (BLSSM-IS), which will serve as the theoretical backdrop for our analysis of TPAs, as it can well embed CPV in its neutrino sector. (For phenomenological studies of the BLSSM, in both Type-I and IS configurations, see Refs.~\cite{Elsayed:2011de,Elsayed:2012ec,Abdallah:2014fra,Hammad:2015eca,Hammad:2016trm,Abdallah:2016vcn,DelleRose:2017hvy,DelleRose:2017ukx,Ahmed:2020lua,Biswas:2018yus,Abdallah:2019svm,Kanemura:2014rpa}.)
 
 The structure of this paper is as follows. The BLSSM-IS model and how CPV signatures can emerge in its leptonic sector is described in Section \ref{sec:CPV}. This is followed by a section introducing some Benchmark Points (BPs) over the parameter spaces of this BSM scenario alongside presenting some details of our Monte Carlo (MC) simulation. In the following part we perform our collider analysis using both a cut-based analysis and a Boosted Decision Tree (BDT) and present the ensuing results, including a discussion of the $Z'$ portal used and of the CP-Conservation (CPC) limit. We then conclude.

%###################################################
\section{The BLSSM-IS scenario and CPV}\label{sec:CPV}

The BLSSM-IS extends the MSSM by adding three parts. The first part contains two SM singlet chiral Higgs superfield $\chi_1$ and $\chi_2$ with $B-L$ charge equal to $+1$ and $-1$, respectively.  When the scalar components of the superfields $\chi_1$ and $\chi_2$ get their VEVs, they break the $U(1)_{B-L}$ symmetry spontaneously. The second part includes three SM singlet chiral superfields: $N_i$ ($i=1,2,3$) with $B-L$ charge equal to $-1$. They are used to cancel the $U(1)_{B-L}$ anomaly and the fermionic component of $N_i$ is considered an RH neutrino. The third part consists of two chiral SM singlet superfield $S_1$ and $S_2$ with a $B-L$ charge equal to $+2$ and $-2$, respectively, to implement the inverse seesaw mechanism \cite{Elsayed:2011de, Khalil:2010iu}. The superpotential of the leptonic sector is
\begin{equation}\label{eq:superpotential}
    W = Y_E L H_1 E^c + Y_\nu L H_2 N^c + Y_S N^c\chi_1 S_2 + \mu H_1 H_2 + \mu^\prime \chi_1 \chi_2 + \mu_{S}S_{2}S_{2}.
\end{equation}
The last one of these can only be an effective term generated through non-renormalizable interactions of the form $\frac{\chi_1^4 S_2^2}{M_I^3}$, where $M_{I}$ is some intermediate mass scale. One can of course add a number of other terms to the superpotential. For instance the terms
\begin{equation}
    W^{\prime} = \mu_{12}S_{1}S_{2}+\mu_{11}S_{1}S_{1}\label{eq:superprime}
\end{equation}
could be added, the latter being generated through non-renormalizable terms like the $S_{2}S_{2}$ term. These terms can be involved in neutrino mass generation, but since Majorana masses are proportional to the lepton number violating terms in the model, the active neutrino masses remain small as long as both $\mu_{11}$ and $\mu_{S}$ are small and the lepton number conserving mass terms are large enough. Here we simply assume that the contribution to neutrino masses from equation (\ref{eq:superprime}) is subdominant compared to equation (\ref{eq:superpotential}), which is the common assumption in this model. That is the case if $\mu_{12}\gg Y_{S}\langle \chi_{1}\rangle$ and $\mu_{11}\ll \mu_{S}$.

When the EW symmetry and the $B-L$ symmetry break, the Higgs fields and the $\chi_i$ ($i=1,2$) get a VEV. This leads to neutrino mass terms
\begin{equation}
    \mathcal{L} = m_D \Bar{\nu}_L N^c + M_N \Bar{N}^c S_2 +\mu_{S} S_{2}S_{2}, 
\end{equation}

where $m_D = Y_\nu v \sin{\beta}$ and $M_N = Y_S v^\prime \sin{\theta}$, with $\tan \beta = \frac{v_2}{v_1} = \frac{\langle H_2 \rangle}{\langle H_1 \rangle}$ and $\tan \theta = \frac{v_1^\prime}{v_2^\prime} = \frac{\langle\chi_1 \rangle}{\langle \chi_2 \rangle}$. Specifically, $v^2 = v_1^2 + v_2^2$ and ${v^\prime}^2 = {v^\prime_1}^2 + {v^\prime_2}^2$. The smallness of Left-Handed (LH) neutrino masses are related to a small lepton number violating term $\mu_S S_2 S_2$, with $\mu_S \approx \mathcal{O}(1)$ keV, which can emerge at the $B-L$ scale from the non-renormalizable term in the superpotential $\frac{\chi_1^4 S_2^2}{M_I^3}$, with $M_I$ an intermediate scale of the order $\mathcal{O}(10^7)$ GeV \cite{Elsayed:2011de}. The $3\times3$ neutrino mass matrix of one generation has the following form (in the basis $(\nu_L, N^c, S_2)$):
\begin{equation}
    M_\nu = \left( \begin{array}{ccc}
        0 & m_D & 0\\
        m_D & 0 & M_N \\
        0 & M_N & \mu_S 
    \end{array}\right).
\end{equation}

one finds that the light neutrino masses are
\begin{equation}
    m_{\nu_L} = \frac{m^2_D \mu_S}{M^2_N + m^2_D}
\end{equation}
and the heavy ones being nearly degenerate with a mass
\begin{equation}
    m = \sqrt{M^2_N + m^2_D}\simeq M_{N}.
\end{equation}

This model is of interest for studying the CPV in the sneutrino sector for three reasons. First, the additional gauge boson $Z^{\prime}$ couples to the sneutrinos and can act as a portal to produce them. Without such a portal the production cross section for heavy sneutrinos would be low even at future colliders. Second, the inverse seesaw mechanism allows for large superpotential couplings in the neutrino-sneutrino sector, which is necessary for the CPV phases to create large effects in observables. In BLSSM-IS both $Y_{\nu}$ and $Y_{S}$ can be large. Their relative importance depends on the detailed decay chain of the sneutrinos. { Third, the breaking of U$(1)_{B-L}$ can provide a first-order phase transition necessary to create a large enough matter-antimatter asymmetry. If the symmetry breaking scale is a few TeV's successful leptogenesis is possible with $\mathcal{O}(1)$ CP-violation \cite{Blanchet:2009bu} as long as the reheating temperature is sufficiently large \cite{Racker:2008hp}. If we were to discover an additional gauge boson, say, at the HL-LHC, it would be important to estimate whether we have new sources of CPV that could account for the observed matter-antimatter asymmetry.}

Next, we show how CPV could be observed in the sneutrino sector. If any of the superpotential coefficients in Equation (\ref{eq:superpotential}) is complex, this can lead to CPV, provided one cannot absorb the phases by redefining the superfields. The Yukawa terms arising from a complex superpotential coupling $y_{ijk}X_{i}X_{j}X_{k}$ are of the form
\begin{equation}
\mathcal{L}=y_{ijk}\bar{\Psi}_{i}\left(\frac{1-\gamma^{5}}{2}\right)\Psi_{j}\varphi_{k}+y_{ijk}^{*}\bar{\Psi}_{i}\left(\frac{1+\gamma^{5}}{2}\right)\Psi_{j}\varphi_{k}^{*}+\mathrm{permutations}.
\end{equation}
As long as the coupling $y_{ijk}$ is neither real nor purely imaginary, there will be both a scalar and a pseudoscalar coupling between the fermions $i,j$ and the scalar $k$ (+permutations in $ijk$). This leads to direct CPV, since the interference of the scalar and pseudoscalar parts produces a parity violating distribution. Since $\bar{\Psi}\Psi$ and $i\bar{\Psi}\gamma^{5}\Psi$ are both even under charge conjugation, conservation of CPT tells us that this part is also T-odd. Hence, triple products of momenta, of the form $p_{1}\cdot (p_{2}\times p_{3})$, which are odd both under P and T will be sensitive to such a form of CPV. 

There is also another source for CPV effects that arises from the complex Yukawa couplings. That comes from the mixing of CP-even and CP-odd sneutrino states in the sneutrino mass matrix induced by the complex Yukawa couplings. The mixture of CP-even and CP-odd components in the amplitude will again lead to terms with varying numbers of $\gamma^{5}$ factors in them. In the general case, the mixing between CP-even and CP-odd states can also be a result of complex soft SUSY breaking mass matrices. Here we take the soft SUSY breaking masses to be real.

\begin{figure}[htb]
\centering
\includegraphics[width=0.6\textwidth]{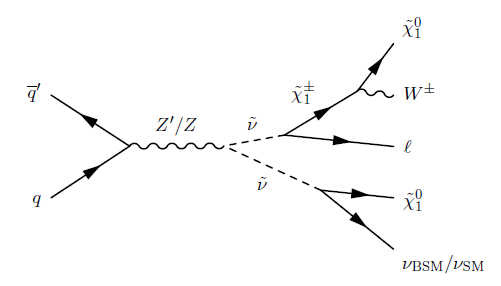}
\caption{The Feynman diagram showing one of the possible production and decay modes of the sneutrinos. The $W^\pm$ will eventually decay hadronically giving multiple jets in the final state. Here all the three flavours of SM neutrinos are collectively denoted as $\nu_{\rm SM}$, while all the six sterile neutrinos are denoted as $\nu_{\rm BSM}$. }
\label{BLSSM-IS}
\end{figure}

We consider the process $p p \rightarrow Z^\prime/Z \rightarrow \Tilde{\nu} \Tilde{\nu}$, when one of the sneutrinos decays into the lightest chargino and a lepton, while the other sneutrino decays to a neutrino and a neutralino. The charginos cascade into the lightest neutralino and $W^\pm$ boson, leading to a di-jet final state when the $W$ decays hadronically\footnote{{We do not consider the leptonic decay of $W^\pm$ boson as the decay will provide us with one visible object, i.e., the lepton. To construct the triple products of our primary interest, we need at least three visible objects. Therefore, even though the leptonic mode is cleaner in the collider environment, we opt for the hadronic mode to enhance the cross section and also to calculate the T-odd observables using the visible objects only, following Eq.~\ref{eq:T-odd}.}}. As a result, we get a final state signature involving di-jets, at least one isolated lepton, and a significantly large MET. The Feynman diagram of such a process is shown in Fig. \ref{BLSSM-IS}. The CPV phases in the (s)neutrino sector can arise from both the $Y_{\nu}$ and $Y_{S}$ couplings of Equation (\ref{eq:superpotential}) and the mixing of the CP-even and CP-odd states.

From the structure of the Superpotential we may notice that direct CPV can give a detectable final state only for RH sneutrinos through the term $Y^{\nu}LH_{2}N^{c}$, while for LH sneutrinos this term would lead to the invisible neutrino+neutralino final state. CPV arising from left-handed sneutrinos comes hence from mixing and only the wino-components of the charginos will contribute. If the wino mass parameter is larger than the sneutrino mass, this contribution is somewhat suppressed.

The most dominant contribution of an SM background here arises from the process: $p p \rightarrow W^\pm + {\rm jets} $, where the $W^\pm$ boson decays into lepton and neutrino. However, processes like $t\bar{t} + {\rm jets}$ with semi-leptonic/di-leptonic final states can also contribute significantly. As we discuss below, we are not going to be looking at the expectation value of T-odd observables, which would limit us to only processes involving the phase of the CKM matrix. We expect that especially the $Z^{\prime}$ mediated signal will contain leptons and jets that are on average harder than what is typical for SM processes. Hence, the contribution from $t\bar{t}$, which tends to lead to hard jets and lepton in the SM, will be an important background.

In this paper, we focus on the CPV effects through the cascaded decays of heavy Supersymmetric particles. When an unstable heavy particle decays to four lighter particles through a cascade, the decay rate includes a term of the form $\epsilon_{\mu\nu\alpha\beta}p_0^\mu p_1^\nu p_2^\alpha p_3^\beta$, where $p_i$ is the four-momentum of the $i$-th particle \cite{Langacker:2007ur}. This arises from the trace of an odd number of $\gamma^{5}$ matrices together with four $\gamma$-matrices. Many such combinations exist in the decay amplitude, \textit{e.g.} taking the vector parts of the $W$-boson vertices together with a $\gamma^{5}$ arising from the pseudoscalar part of the complex Yukawa coupling. In the rest frame of the decaying particle, this term provides the CP-odd observable given by the expectation value of the triple product,
\begin{equation}
    \mathcal{T}=\Vec{p_1}\cdot(\Vec{p_2}\times \Vec{p_3}) \label{eq:T-odd}
\end{equation}
This T-odd observable (triple product) is sensitive to the CP-odd phases and provides a good discriminator between the signal and background (having only complex phases in the CKM matrix as the CP-odd phase). We further formulate the T-odd observables depending on the final state topology of the signal/background.

%\begin{figure}[h!]
%    \centering
%    \begin{fmffile}{BLSSM-IS}
%  \begin{fmfgraph*}(200,120)
%    \fmfstraight
%   \fmfleft{i1,i2,i3,i4,i5}
%   \fmfright{o1,o2,o3,o4,o5}
%   \fmf{fermion, tension=3}{v1,i4}
%   \fmf{fermion, tension=3}{i2,v1}
%   \fmflabel{$q$}{i2}
%   \fmflabel{$\overline{q}^{\prime}$}{i4}
%    \fmf{phantom, tension=0.5}{i5,v2,o5}
%   \fmf{boson, tension=5, label={$Z^\prime$/$Z$}}{v1,v2}
%   \fmf{dashes, label={$\tilde{\nu}$}, tension=3, label.side=right}{v4,v2}
%   \fmf{dashes, label={$\tilde\nu$}, tension=3, label.side=left}{v3,v2}
%   \fmf{fermion, tension=2}{v3,o2}
%   \fmf{fermion, tension=2}{v3,o1}
%   \fmf{fermion, label={$\tilde{\chi}_1^\pm$}, label.side=left}{v4,v5}
%   \fmf{fermion}{v4,o3}
%   \fmf{fermion}{v5,o5}
%   \fmf{boson}{v5,o4}
%   \fmflabel{$\tilde{\chi}_1^0$}{o5}
%   \fmflabel{$W^\pm$}{o4}
%   \fmflabel{$\ell$}{o3}
%   \fmflabel{$\nu_{\rm BSM}/\nu_{\rm SM}$}{o1}
%   \fmflabel{$\tilde{\chi}_1^0$}{o2}  
%  \end{fmfgraph*}
%    \end{fmffile}
%    \vspace{0.5cm}

Our three momentum vectors will quite often involve one lepton and two jets. To properly evaluate the triple product of Equation (\ref{eq:T-odd}), we should be able to distinguish a quark jet from an antiquark jet. We shall not try this. Therefore, since at random we pick equal amounts of quark and antiquark jets for a given $p_{i}$, the expectation value $\langle \mathcal{T}\rangle$ will vanish.
However, the non-zero expectation value has an influence on the distribution. We assume that there is an intrinsic variance to the measurement of $\mathcal{T}$, given by
\begin{equation}
    V(\mathcal{T}) = \langle \mathcal{T}^{2}\rangle - \langle \mathcal{T}\rangle^{2},
\end{equation}
when the momenta are in some given order and that this does not depend on $\langle \mathcal{T}\rangle$. 
Then, we assume that there is some distribution $f(\mathcal{T})$ that gives the probability density of $\mathcal{T}$, when the momenta are in some particular order, say, lepton--quark--antiquark. Then $f(-\mathcal{T})$ gives the probability density for the ordering lepton--antiquark--quark and the overall distribution is given by $\frac{1}{2}(f(\mathcal{T})+f(-\mathcal{T}))$. Now,
\begin{eqnarray}
\langle \mathcal{T} \rangle & = & \int_{-\infty}^{\infty}f(\mathcal{T})\mathcal{T}\, \mathrm{d}\mathcal{T},\\
\langle \mathcal{T}^{2} \rangle & = & \int_{-\infty}^{\infty}f(\mathcal{T})\mathcal{T}^{2}\, \mathrm{d}\mathcal{T}.
\end{eqnarray}
The variance for the distribution $\frac{1}{2}(f(\mathcal{T})+f(-\mathcal{T}))$ can be computed from
\begin{eqnarray}
    \langle \mathcal{T}_{\rm sym} \rangle & = & \int_{-\infty}^{\infty}\frac{1}{2}(f(\mathcal{T})+f(-\mathcal{T}))\mathcal{T}\, \mathrm{d}\mathcal{T}=0,\\
    \langle \mathcal{T}_{\rm sym}^{2}\rangle & = & \int_{-\infty}^{\infty}\frac{1}{2}(f(\mathcal{T})+f(-\mathcal{T}))\mathcal{T}^{2}\, \mathrm{d}\mathcal{T}=\langle \mathcal{T}^{2} \rangle= V(\mathcal{T})+\langle\mathcal{T}\rangle^{2},
\end{eqnarray}
where a change of variables $\mathcal{T}\rightarrow -\mathcal{T}$ has been used in the second terms. If we assume that $V(\mathcal{T})$ is independent of $\langle \mathcal{T}\rangle$, then $\langle \mathcal{T}^{2} \rangle$ is larger if $\mathcal{T}$ has a non-zero expectation value. Since CPV leads to a non-zero expecation value, we expect the variance of $\mathcal{T}_{\rm sym}$ to be larger than for a distribution without an expectation value, \textit{i.e.}, the distribution to be wider in the presence of CPV. Hence, we use this as our indicator of CPV. If one can distinguish between quark and antiquark jets, the simpler criterium $\langle \mathcal{T}\rangle \neq 0$ is sufficient.

%###############################################
\section{The BPs and simulation details} 

We wish to study the leptonic CPV in the sneutrino sector. We introduce explicit CPV in terms of two complex Yukawa matrices $Y_{\nu}$ and $Y_{S}$. Of these, the matrix $Y_{\nu}$ is relevant for decays of type $\tilde{\nu}\rightarrow \ell^{\pm}\tilde{H}^{\mp}$. We prepare our particle spectrum in such a way that there is a gaugino-dominated neutralino state sufficiently lighter than the higgsino-dominated states, so that one gets relatively energetic jets and a reasonably large amount of missing momentum from the decay of the higgsino-dominated state. As discussed above, for left-handed sneutrinos only the wino-component of the chargino is relevant, and the CPV effects arise from mixing.

For a detailed analysis we choose the three pairs of BPs given in Tables \ref{tab:input-params}--\ref{tab:mass-spectrum}. Within a pair we take a CPV BP, where Yukawa matrices are complex and a CPC one, where the Yukawa couplings are the real parts of the corresponding CPV couplings and slight adjustments are made to soft SUSY breaking masses so that the mass spectrums are as identical as possible. These pairs allow us to compare the widths of the triple product distributions and to show that the width of the distribution is different, when sizable CPV is involved.

We choose the model parameters in such a manner that the masses of some of the lighter sneutrinos as well as of the neutralinos and charginos are below a TeV, while the rest of the particle spectrum is decoupled to a few TeV (see Tab. \ref{tab:mass-spectrum}). The masses of the $Z^{\prime}$ (Tab.  \ref{tab:input-params}) are chosen slightly above current experimental bounds since, as we discuss in section \ref{sec:Zprime}, the cross section from the SM $Z$ portal alone is too low. The model files are generated using \textsc{Sarah v4.14} \cite{Staub:2015kfa}, which also provides source codes for \textsc{SPheno v4.0.4} \cite{Porod:2003um,Porod:2011nf} through which we obtain the particle spectrum and different couplings relevant to these BPs. 
%
%-----------------------------------------------------
\begin{table}[htb!]
    \centering
    \begin{tabular}{|c|c|c|c|}
    \hline
     Parameter & BP1 & BP2 & BP3 \\ \hline \hline
     $\tan \beta$ & 20 & 10 & 20 \\ \hline
     $\tan \theta$ & 1.05 & 1.08 & 1.06 \\ \hline
     $M_{Z^\prime}$ [TeV] & 5.20 & 5.30 & 5.30 \\ \hline
     \end{tabular}
     \caption{The values chosen for some of the free parameters of the BLSSM-IS.Furthermore, we choose $g_{B-L}=0.46$ for all points.}
    \label{tab:input-params}
\end{table}
%-----------------------------------------------------

We show some general model parameters in the Tab. \ref{tab:input-params} and the CPV Yukawa couplings in Tab. \ref{tab:BPs}. For BP1 we choose relatively small phases, BP2 has large phases, and BP3 intermediate phases. Kinematically, BP1 and BP3 resemble each other, while the neutralino/chargino spectrum is different for BP2 leading to harder leptons and jets. Since our CPV enters only through leptonic interactions, the contribution to hadronic CPV (which is experimentally constrained) comes at higher loop orders. We expect that the contribution to be so small that the experimental constraints are satisfied even with large CPV phases in the leptonic sector.

%  Note that, too large values of the CPV phases will introduce several theoretical and phenomenological problems that are not relevant to the theme of this work. {\textcolor{red}{[AC to Harri: I have added this line to highlight why we can't take very large values of the phases. Please revise/modify the previous sentence if needed.]}} {{[SM to Amit: I am not sure what this sentence means, we would like to have large phases, point is that we cannot because we fail CPV observables, so we should list the latter and mention limits, while making clear that our BPs are not violating these.]}} 

%-------------------------------------
\begin{table}[htb!]
    \centering     
     \begin{tabular}{|c|c|c|c|}
     \hline
     Parameter & BP1   & BP2 & BP3 \\ \hline  \hline
     $Y_{\nu 11}$ & 0.22+0.006i & 0.25+0.14i & 0.34+0.08i    \\ \hline
     $Y_{\nu 12}$ & 0.12+0.0027i & -0.16+0.11i & -0.23+0.067i   \\ \hline
     $Y_{\nu 13}$ & -0.067-0.004i & 0.06-0.04i & 0.086-0.016i \\ \hline
     $Y_{\nu 21}$ & 0.12+0.013i & 0.15+0.01i & 0.25+0.1i   \\ \hline
     $Y_{\nu 22}$ & 0.31+0.011i & 0.43+0.26i & 0.53+0.061i    \\ \hline
     $Y_{\nu 23}$ & 0.17-0.006i & 0.39-0.2i & 0.46-0.1i   \\ \hline
     $Y_{\nu 31}$ & -0.074+0.012i & -0.074-0.046i & -0.094+0.027i  \\ \hline
     $Y_{\nu 32}$ & 0.16+0.053i & 0.17+0.14i & 0.22+0.083i   \\ \hline
     $Y_{\nu 33}$ & 0.34+0.014i & 0.37+0.3i & 0.43+0.12i   \\ \hline
    \end{tabular}   
    \caption{The complex Yukawa matrix elements are shown for the three CPV BPs. For the first one the phases are small, the two others have large phases.}
    \label{tab:BPs}
\end{table}  

\begin{table}[h!]
    \centering
    \begin{tabular}{|c|c|c|c|} \hline
      Particle & BP1 & BP2 & BP3 \\ \hline
      $\tilde{\nu}_1$ & 871.3 & 902.0 & 884.6\\ \hline
      $\tilde{\nu}_2$ & 871.3 & 902.0 & 884.6\\ \hline
      $\tilde{\nu}_3$ & 922.5 & 903.0 & 925.2\\ \hline
      $\tilde{\nu}_4$ & 922.5 & 903.0 & 925.2\\ \hline
      $\tilde{\nu}_5$ & 940.5 & 904.4 & 951.0\\ \hline
      $\tilde{\nu}_6$ & 940.5 & 904.4 & 951.0\\ \hline
      $\tilde{\chi}_1^\pm$ & 583.7 & 659.2 & 583.0\\ \hline
      $\tilde{\chi}_1^0$ & 533.6 & 514.6 & 531.4\\ \hline
      $\tilde{\chi}_2^0$ & 587.1 & 664.2 & 586.5\\ \hline
      $\tilde{\chi}_3^0$ & 602.8 & 664.3 & 602.1 \\ \hline
      $\nu_{4}$ & 16.26 & 31.05 & 35.25 \\ \hline 
      $\nu_{5}$ & 16.26 & 31.05 & 35.25 \\ \hline
      $\nu_{6}$ & 56.30 & 64.34 & 76.47 \\ \hline
      $\nu_{7}$ & 56.30 & 64.34 & 76.47 \\ \hline
      $\nu_{8}$ & 82.20 & 127.6 & 140.2 \\ \hline
      $\nu_{9}$ & 82.20 & 127.6 & 140.2 \\ \hline
    \end{tabular}
    \caption{{The masses (in GeV) of some of the sneutrinos, lighter gauginos, and the six sterile neutrinos ($\nu_4$ to $\nu_9$) for all the three BPs.}}
    \label{tab:mass-spectrum}
\end{table}

The particle spectrum for the three BPs is shown in Tab. \ref{tab:mass-spectrum}. In addition to the masses, the Branching Ratios (BRs) are also modified because of the presence of the CPV phases. In Tab.  \ref{tab:BR}, we display the BRs of the six lightest sneutrinos, namely $\tilde{\nu}_{i}$ ($i$ = 1,..,6), to different possible final states, including the new decay modes originating from BSM physics. The decay modes primarily include charginos and neutralinos along with SM neutrinos and sterile neutrinos. For the leptonic decay of the sneutrinos through charginos, we combine the contributions coming from electrons, muons, and leptonically decaying taus. When a sneutrino decays invisibly, we separately add the contribution from the three SM neutrinos flavors, collectively denoted as $\nu_{\rm SM}$, and contributions from the six sterile neutrinos, collectively denoted as $\nu_{\rm BSM}$. The BRs of other modes that involve neutralinos other than the lightest one are summed for a given sneutrino and are expressed as $BR(\tilde{\nu} \to {\rm others})$. It is evident from the table that the presence of imaginary parts in the couplings affects the BRs (both visible and invisible ones), especially in the case of the `BP2' where the imaginary parts are relatively large compared to the other two BPs. The complex couplings and the mixing of CP eigenstates also affect the angular distributions of the decay products, which in turn will affect the distribution of T-odd observables.

%-------------------------------------
\begin{table}[h!]
    \centering
    \begin{tabular}{|c|c|c|c|}\hline
      Decay channels   & BP1 & BP2 & BP3 \\ \hline
      BR($\tilde{\nu}_1 \to \tilde{\chi}_1^\pm  l^\mp$)  &0.0251& 0.4912  & 0.0346 \\ 
      BR($\tilde{\nu}_1 \to \tilde{\chi}_1^0  \nu_{\rm SM}$)  &0.2068&0.0018&  0.2796  \\
      BR($\tilde{\nu}_1 \to \tilde{\chi}_1^0  \nu_{\rm BSM}$)  &0.2931&0.4817&  0.3570 \\ 
      BR($\tilde{\nu}_1 \to {\rm others}$)  & 0.474 & 0.026 &  0.332\\ \hline
      
      BR($\tilde{\nu}_2 \to \tilde{\chi}_1^\pm  l^\mp$)  &0.0251 &0.4912& 0.0346 \\ 
      BR($\tilde{\nu}_2 \to \tilde{\chi}_1^0  \nu_{\rm SM}$)  &0.2086 &0.0018&0.2796    \\
      BR($\tilde{\nu}_2 \to \tilde{\chi}_1^0  \nu_{\rm BSM}$)  &0.2939&0.4808& 0.3568  \\ 
      BR($\tilde{\nu}_2 \to {\rm others}$)  & 0.474 & 0.026 &  0.332\\ \hline   
      
      BR($\tilde{\nu}_3 \to \tilde{\chi}_1^\pm  l^\mp$)  &0.0331&0.4890& 0.0368  \\ 
      BR($\tilde{\nu}_3 \to \tilde{\chi}_1^0  \nu_{\rm SM}$)  &0.072&0.0001& 0.2921  \\
      BR($\tilde{\nu}_3 \to \tilde{\chi}_1^0  \nu_{\rm BSM}$)  &0.3095&0.0044&0.3988    \\ 
      BR($\tilde{\nu}_3 \to {\rm others}$)  & 0.584  & 0.496 &  0.263\\ \hline
      
      BR($\tilde{\nu}_4 \to \tilde{\chi}_1^\pm  l^\mp$)  &0.0311& 0.4890 &0.0368  \\ 
      BR($\tilde{\nu}_4 \to \tilde{\chi}_1^0  \nu_{\rm SM}$)  &0.072& 0.0001 & 0.2922   \\
      BR($\tilde{\nu}_4 \to \tilde{\chi}_1^0  \nu_{\rm BSM}$)  &0.3091& 0.0044 & 0.3985  \\ 
      BR($\tilde{\nu}_4\to {\rm others}$)  & 0.584 & 0.496 & 0.263 \\ \hline  
      
      BR($\tilde{\nu}_5 \to \tilde{\chi}_1^\pm  l^\mp$)  &0.1933 &0.4993&0.318  \\ 
      BR($\tilde{\nu}_5 \to \tilde{\chi}_1^0  \nu_{\rm SM}$)  &0.0258&--&  0.0403 \\
      BR($\tilde{\nu}_5 \to \tilde{\chi}_1^0  \nu_{\rm BSM}$)  &0.2559&0.026& 0.4024   \\ 
      BR($\tilde{\nu}_5 \to {\rm others}$)  & 0.521 & 0.472 & 0.231 \\ \hline
      
      BR($\tilde{\nu}_6 \to \tilde{\chi}_1^\pm  l^\mp$)  &0.1933& 0.4993 &0.318  \\ 
      BR($\tilde{\nu}_6 \to \tilde{\chi}_1^0  \nu_{\rm SM}$)  &0.0258&--&  0.0403 \\
      BR($\tilde{\nu}_6 \to \tilde{\chi}_1^0  \nu_{\rm BSM}$)  &0.2559&0.026&  0.4009 \\ 
      BR($\tilde{\nu}_6 \to {\rm others}$)  & 0.521 & 0.472 & 0.231 \\ \hline      
    \end{tabular}
    \caption{The BR of $\tilde{\nu}_{i}$ ($i$ = 1,..,6) to different possible final states, including the new decay modes originating from the BSM physics, corresponding to BP1, BP2 and BP3. See the text for more details.}
    \label{tab:BR}
\end{table}
%--------------------------------------
%\subsection{Simulation details}

The SM background processes, e.g., $pp\to W^{\pm} + {\rm jets}$, $t\bar{t} + {\rm jets}$ etc., will receive the CPV effects through the phase present in the CKM matrix. However, as discussed in section \ref{sec:CPV}, we are looking at the overall width of the distribution and will be insensitive to the deviation from zero, which is the traditional measure of CPV. Hence, it is important to include the SM processes, which lead to largest values of our triple products. It is also critical that the modeling of the CPC BSM physics is accurate enough as we need to show a wider distribution than would emerge from the sneutrino processes without CPV.

Therefore, as already mentioned, we will calculate several observables that are sensitive to the CPV effects. To do so, we simulate 200,000 signal events and around 500,000 background events, both at Leading Order (LO) in $\alpha_s$, using \textsc{Madgraph5 (v2.8.2)} \cite{Alwall:2011uj} using collision energy $\sqrt s$ = 100 TeV and integrated luminosity $\mathcal{L} = 3 ~ \rm ab^{-1}$ at the FCC-hh collider \cite{Benedikt:2022kan}. These events are then passed on to \textsc{Pythia (v8.2)} \cite{Sjostrand:2014zea} for parton showering and hadronization, plus finally to \textsc{Delphes3} \cite{deFavereau:2013fsa} to consider the detector effects. The Delphes card in the FCC-hh is redesigned to satisfy higher Pile-Up (PU) and enhance the tagging efficiencies. To efficiently manage the simulated background events, we apply a generation level cut on the minimum missing transverse momentum calculated using the invisible particles present at the parton level setting at 100 GeV. This has helped to generate events within the desired signal region of interest, especially regions with large missing transverse energy. The cross sections for the signal and background processes are listed in Tab. \ref{tab:BP-cross section}. 
The jets are reconstructed using {\texttt FastJet} \cite{fastjet} with jet radius R = 0.4 and anti-$k_T$ jet algorithm \cite{Cacciari:2008gp}. The jets are pre-selected with the jet transverse momentum $p_T^j>30~\rm GeV$ and the pseudorapidity $|\eta|<4.5$.

%-----------------------

\begin{table}[htb!]
    \centering
    \begin{tabular}{|c|c|c|}
    \hline
    & Process & Cross section ($\sigma$) [pb]  \\ \hline 
     &$ p p \to \tilde{\nu}_{i} \tilde{\nu}_{j}$ (BP1) & $5.2\times 10^{-3}$ \\  
     \cline{2-3}
    Signals &$ p p \to \tilde{\nu}_{i} \tilde{\nu}_{j}$ (BP2) & $6.4\times 10^{-3}$\\  
     \cline{2-3}
    &$ p p \to \tilde{\nu}_{i} \tilde{\nu}_{j}$ (BP3) & $5.3\times 10^{-3}$ \\  
     \hline
     &$W^{\pm} + {\rm jets} $  & $2.93\times 10^3$ \\ 
     \cline{2-3}
     &$t\Bar{t}~(1L)$ & $1.49 \times 10^3$ \\ \cline{2-3}
   Backgrounds  &$t\Bar{t}~(2L)$ & $3.42 \times 10^2$ \\ \cline{2-3}
     &$t W$ & 1.37 \\ \cline{2-3}
     &$t\Bar{t}W$ & 0.518 \\ \hline
    \end{tabular}
    \caption{The LO cross sections of the simulated signal and background processes with $\sqrt{s}=100 ~\rm TeV$ at FCC-hh.}
    \label{tab:BP-cross section}
\end{table}

%%%%%%%%%%%%%%%%%%%%%
%\input{collider}

\section{Collider phenomenology and results}

In this section, we discuss the details of the collider analysis, including the multiVariate analysis, followed by the results obtained to probe the sensitivity of the proposed 100 TeV collider experiment.
%%%%%%%%%%%%%%%%%%%%%%%%%%%%%%%%%%%
\subsection{Kinematic variables}

\begin{figure}[htb!]
    \centering
    \includegraphics[scale=0.3]{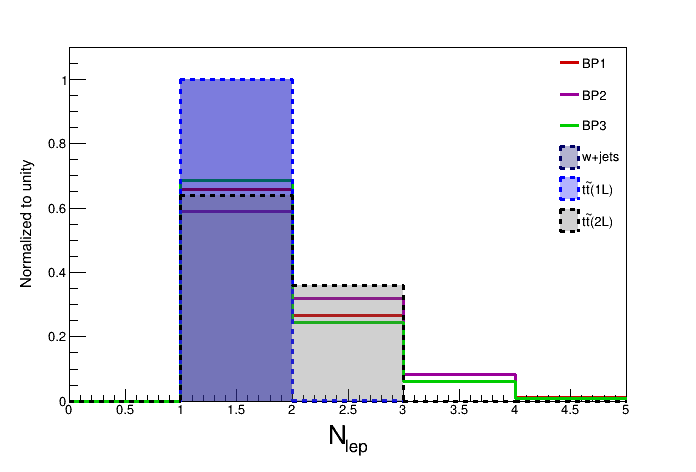}
    \includegraphics[scale=0.3]{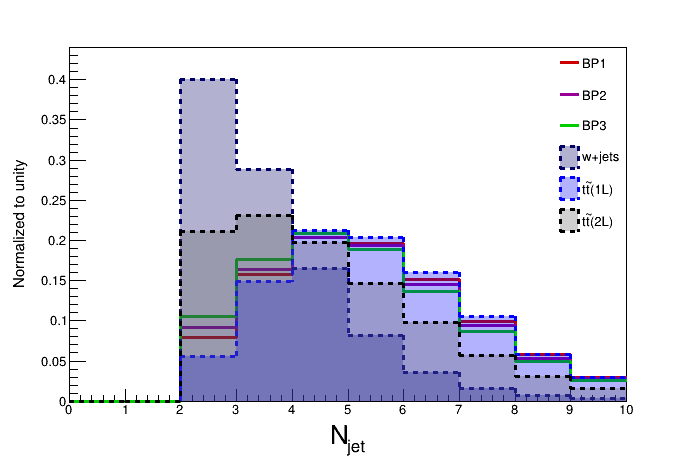}
    \caption{Distribution of the number of leptons (left) and jets (right) for the signals and SM background.}
    \label{fig:njet}
\end{figure}

\begin{figure}[htb!]
\centering
\begin{subfigure}{0.33\textwidth}
  \centering
  \includegraphics[width=1\linewidth]{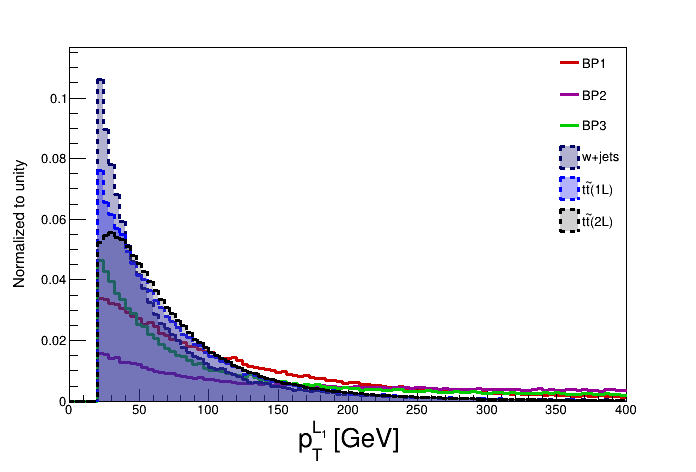}
\end{subfigure}%
\begin{subfigure}{.33\textwidth}
  \centering
  \includegraphics[width=1\linewidth]{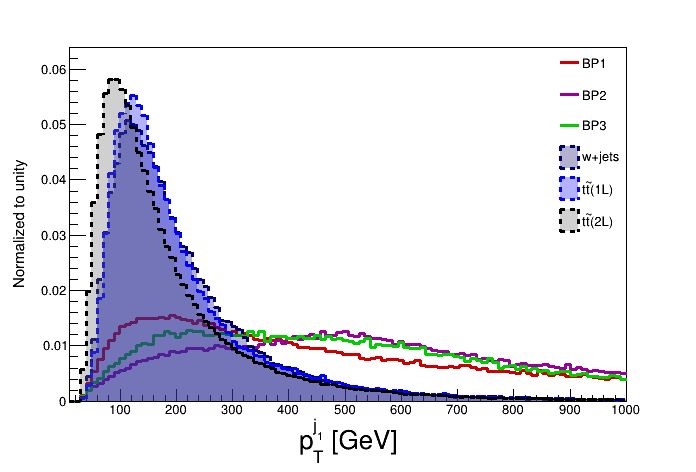}
\end{subfigure}%
\begin{subfigure}{.33\textwidth}
  \centering
  \includegraphics[width=1\linewidth]{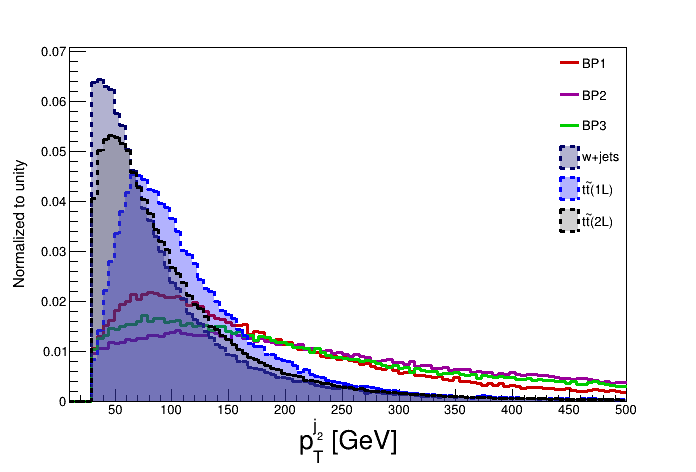}
\end{subfigure}%
\caption{{The left panel describes the transverse momentum of the leading lepton. The middle panel shows the distribution of transverse momentum for the leading jet ($p^{j_1}_T$). The right panel presents the $p_T$ of the sub-leading jet ($p^{j_2}_T$).}}
\label{fig:jetPT}
\end{figure}

As already mentioned, we focus on the final state topology that includes at least one isolated lepton ($e$ or $\mu$), at least two jets along with large MET, say ${\mathcal {O}} (100)$ GeV. These leptons must satisfy $p_T^L>20~\rm GeV$ with pseudorapidity $|\eta|<2.5$. In Fig. \ref{fig:njet}, we show the multiplicity of leptons (left) and jets (right) present in the signal and background events\footnote{In the subsequent analysis we consider the three dominant backgrounds only, neglecting the two sub-dominant ones, namely, $tW$ and $t\bar{t}W$, as their contribution to the forthcoming classification step is highly negligible.}. The signal events include significantly more leptons and jets compared to the background events. We check that these leptons originated primarily from the decay of the sneutrinos to a sterile neutrino and a neutralino, where these neutrinos further decay to leptons and jets. The mass difference between these new particles ensures that some of the lepton/jets are relatively boosted. This is evident from the $p_T$ distribution of the leading (i.e., with highest $p_T$) lepton and leading two jets as shown in Fig. \ref{fig:jetPT}. The sub-leading jet seems to overlap with the SM backgrounds and, therefore, will not play a significant role in controlling the latter. 

%--------------------------------------
%--------------------------------------
%--------------------------------------
%---------------------------

\begin{figure}[htb!]
    \centering
\begin{subfigure}{0.33\textwidth}
  \centering
  \includegraphics[width=1\linewidth]{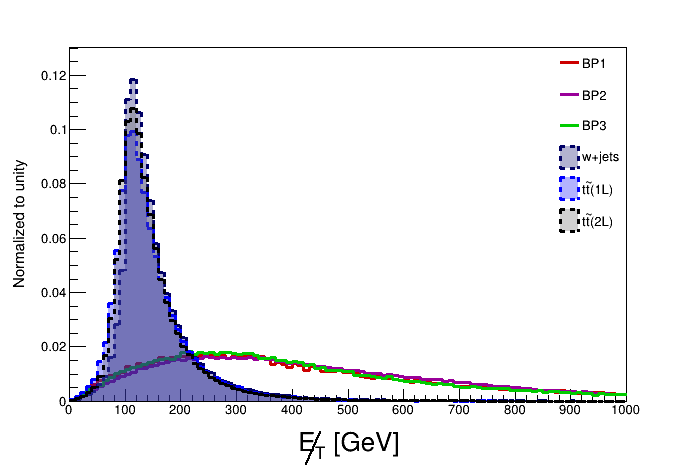}
\end{subfigure}%
\begin{subfigure}{0.33\textwidth}
  \centering
  \includegraphics[width=1\linewidth]{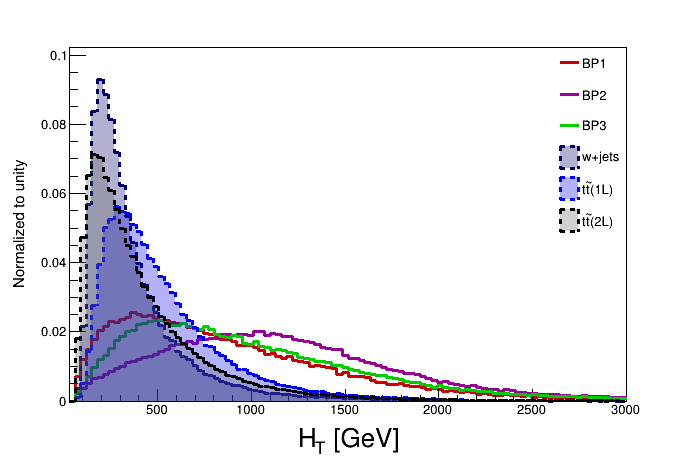}
\end{subfigure}%
\begin{subfigure}{0.33\textwidth}
  \centering
  \includegraphics[width=1\linewidth]{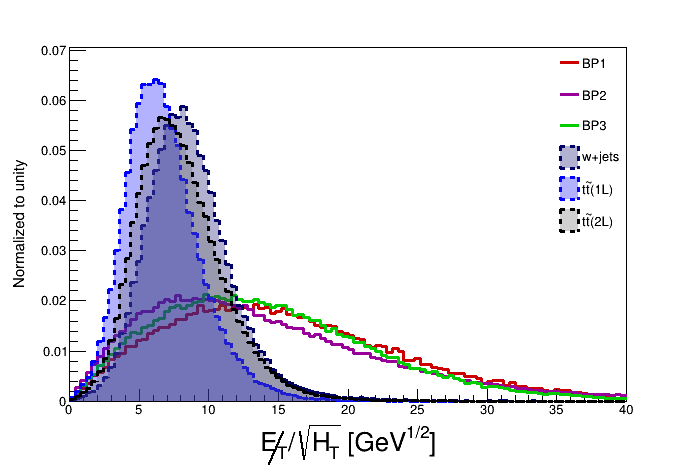}
\end{subfigure}%
    \caption{{Distribution in MET ($\slashed{E}_{T}$) for the signal and SM backgrounds (left). Distribution in the $H_T$ variable describing the scalar $p_T$ sum of all the jets present in that event (middle). Distribution in the ratio  $\slashed{E}_{T}/\sqrt{H_T}$ (right).}}
    \label{fig:pTL-ETMiss}
\end{figure}

The presence of additional invisible particles, in addition to the SM neutrinos, and the contribution of mistagging of the reconstructed objects leads to higher MET values (denoted by $\slashed{E}_{T}$ in the plots) for signal events and, therefore, provides a clear separation from the SM backgrounds. The left panel of Fig. \ref{fig:pTL-ETMiss} shows the MET distribution. It seems that a choice of selecting events with $\slashed{E}_{T}>300~\rm GeV$ would allow most of the signal to be retained while discarding a significant fraction of the backgrounds. Additionally, we also construct two more important variables, namely $H_T$ and $M_{\rm eff}$, where $H_T$ denotes the scalar $p_T$ sum of all jets present in the event, while $M_{\rm eff}$ includes all  visible (lepton and jet) and invisible objects transverse momentum. In the middle panel of Fig. \ref{fig:pTL-ETMiss}, we show the distribution of $H_T$ while in the right panel we display the distribution of the ratio $\frac{\slashed{E}_{T}}{\sqrt{H_T}}$. Both of these distributions show a clear separation power of the signal with respect to the backgrounds.  
%--------------------------------------
%-----------------------------------

\begin{figure}[!htb]
    \centering
    \includegraphics[width=0.4\linewidth]{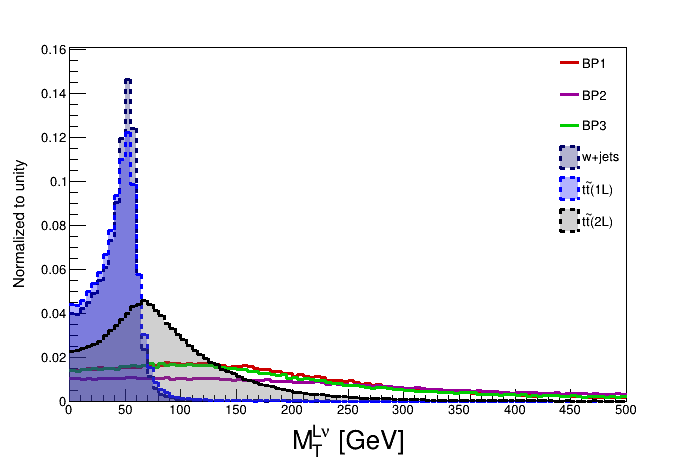}
    \includegraphics[width=0.4\linewidth]{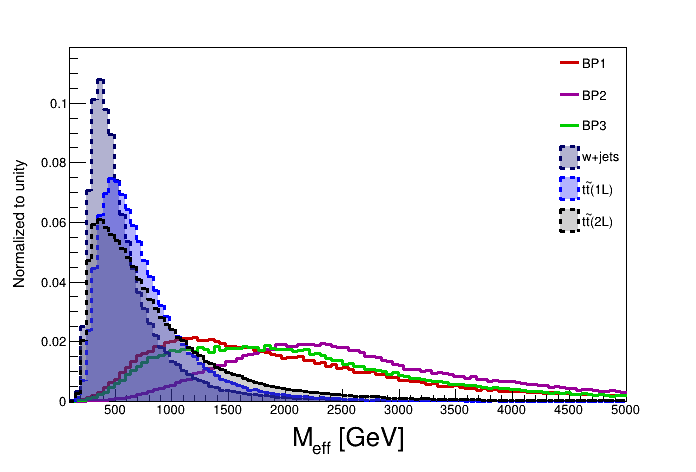}
    \caption{Left: distribution in the transverse mass of the leading lepton and MET. Right: distribution in $M_{\rm eff}$. We refer to the text for more details.}
    \label{fig:MT}
\end{figure}

\begin{figure}[htb!]
    \centering
    \includegraphics[width=0.4\linewidth]{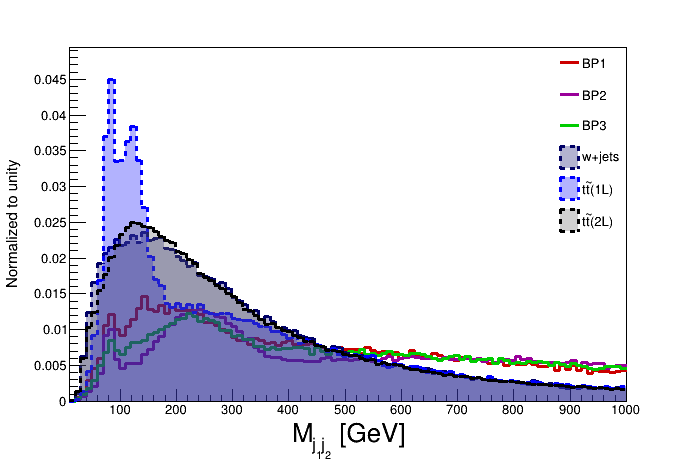}
    \includegraphics[width=0.4\linewidth]{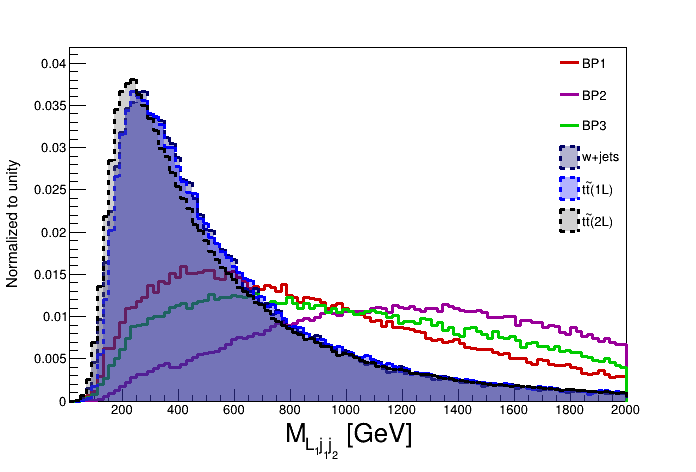}
    \caption{{Left: distribution in the invariant mass of the two leading jets. Right: distribution in the invariant mass of the leading lepton and the two leading jets.}}
    \label{fig:Inv-mass}
\end{figure}

In the left panel of Fig. \ref{fig:MT}, we show the distribution of the transverse mass $M_T$ constructed using the leading lepton and $\slashed{E}_{T}$, defined as $M_T^{L\nu}=\sqrt{p_T^{L}\slashed{E}_{T}(1-\cos(\Delta \phi_{L\slashed{E}_{T}}))}$, where $\phi$ is the azimuthal angle between the lepton momentum and $\slashed{E}_{T}$. For the SM backgrounds, the primary source of this lepton and MET is a $W$-boson that decays leptonically. Therefore, the $M_T$ distribution is expected to have an edge around the $W$ mass (i.e. 80 GeV), which can be seen from the distribution. The signal events arise from the decay of a heavier object that decays to more than just a lepton and a MET (in the form of a neutralino), so there will be no edges and the distribution extends to higher values of $M_{T}$, as seen from the figure. Note that for the $t\bar{t} (2L)$ background, which implies that both the W-bosons are decaying leptonically, the same distribution gets some shift towards higher values. This can be understood from the fact that these events have more than one lepton to choose from the set and also have somewhat higher MET. The variable $M_{\rm eff}$ also takes into account the effect of large MET, and so has a clear separation from the SM backgrounds, as shown in the right panel of Fig, \ref{fig:MT}. Along with these observables, we also calculate the invariant mass of the two leading jets and the same using the leading lepton and the two leading jets, as shown in Fig. \ref{fig:Inv-mass}. 

%--------------------------------------
%-----------------------------------
%--------------------------------------
%----------------------------------

%--------------------------------------

\begin{figure}[htb!]
\centering
\begin{subfigure}{0.33\textwidth}
  \centering
  \includegraphics[width=1\linewidth]{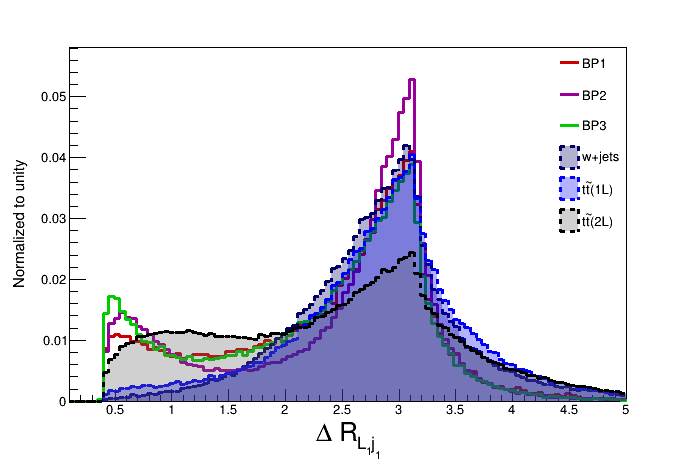}
\end{subfigure}%
\begin{subfigure}{.33\textwidth}
  \centering
  \includegraphics[width=1\linewidth]{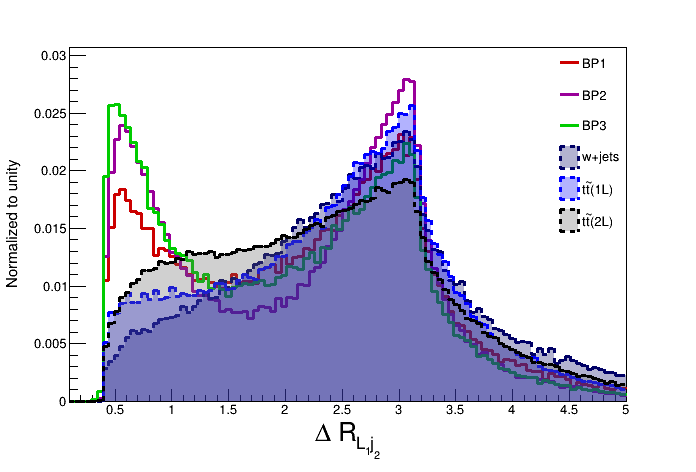}
\end{subfigure}%
\begin{subfigure}{.33\textwidth}
  \centering
  \includegraphics[width=1\linewidth]{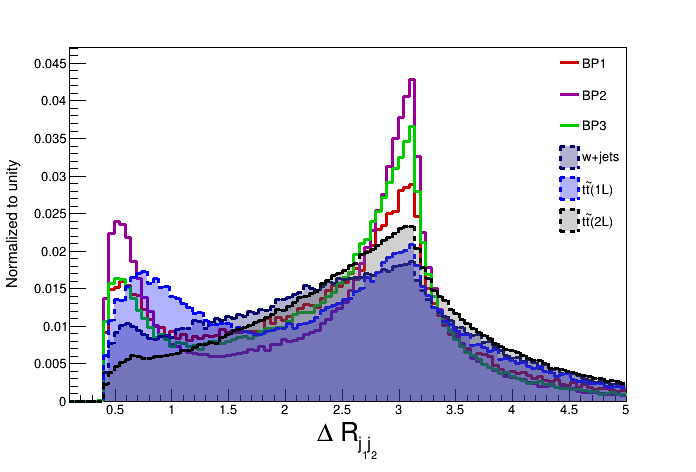}
\end{subfigure}%
\caption{$\Delta R$ distribution between the lepton and leading jet (left), the lepton and second leading jet (middle) as well as the first and second leading  jet (right).}
\label{fig:Delta-r}
\end{figure}

Having discussed some of the kinematic variables involving the lepton and jets, we also need to assess how the leptons and jets are separated among each other. We calculate the angular separation $\Delta R$ between the leading lepton with the two leading jets, where $(\Delta R)^2 = (\Delta \phi)^2 + (\Delta \eta)^2$ in the ($\eta , \phi$) plane, where $\phi$ is the azimuthal angle and $\eta$ is the pseudorapidity of the individual objects. In Fig. \ref{fig:Delta-r}, we show the $\Delta R$ distributions between the leading lepton and leading jet (left panel), leading lepton and sub-leading jet (middle panel), and the two leading jets (right panel). 

\begin{figure}[htb!]
\centering
\begin{subfigure}{0.33\textwidth}
  \centering
  \includegraphics[width=1\linewidth]{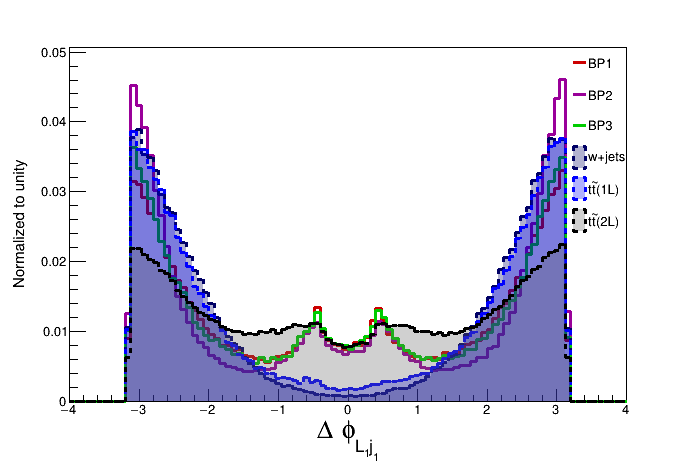}
\end{subfigure}%
\begin{subfigure}{.33\textwidth}
  \centering
  \includegraphics[width=1\linewidth]{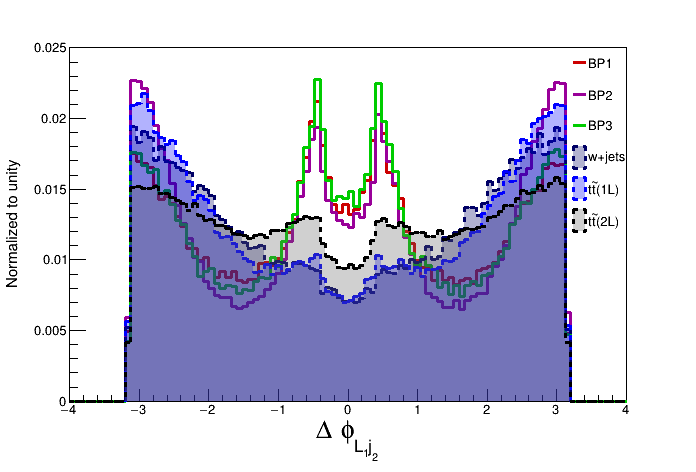}
\end{subfigure}%
\begin{subfigure}{.33\textwidth}
  \centering
  \includegraphics[width=1\linewidth]{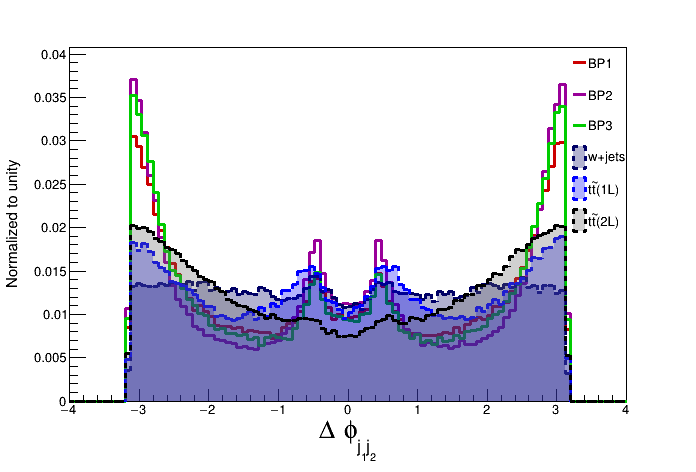}
\end{subfigure}%
\caption{$\Delta \phi$ distribution between the lepton and leading  jet (left), the lepton and second leading  jet (middle) as well as the first and second leading  jet (right).}
\label{fig:Delta-phi}
\end{figure}

\begin{figure}[htb!]
\centering
\begin{subfigure}{0.33\textwidth}
  \centering
  \includegraphics[width=1\linewidth]{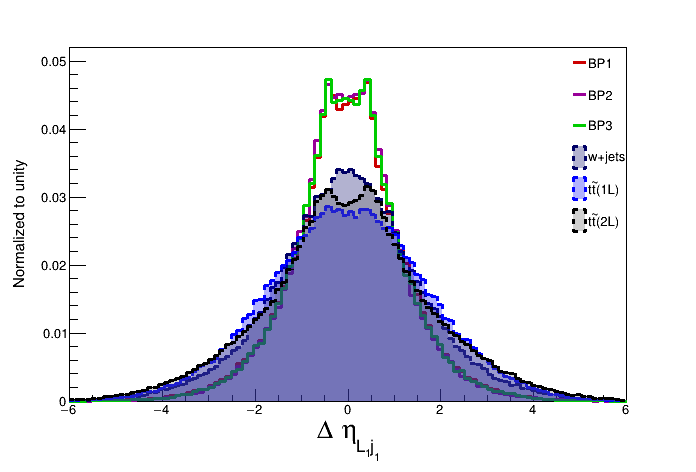}
\end{subfigure}%
\begin{subfigure}{.33\textwidth}
  \centering
  \includegraphics[width=1\linewidth]{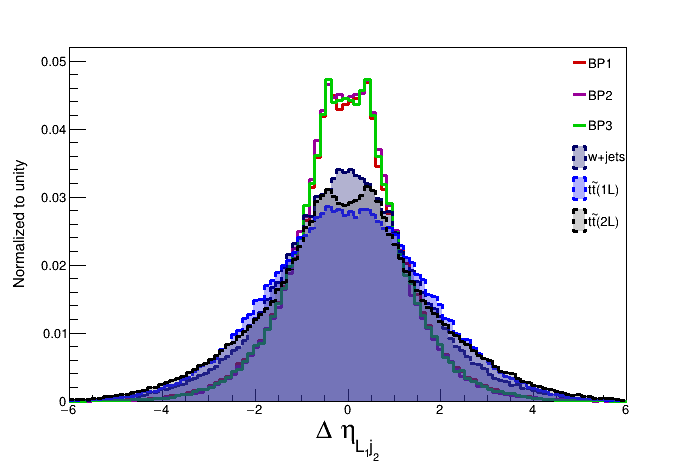}
\end{subfigure}%
\begin{subfigure}{.33\textwidth}
  \centering
  \includegraphics[width=1\linewidth]{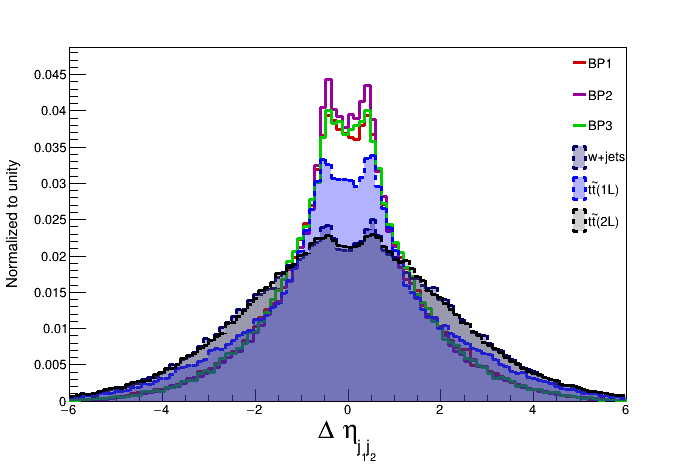}
\end{subfigure}%
\caption{$\Delta \eta$ distribution between the lepton and leading  jet (left), the lepton and second  leading jet (middle) as well as the first and second leading  jet (right).}
\label{fig:Delta-eta}
\end{figure}

The leading lepton seems to be more aligned with the direction of the leading jet, while it can be either very close or widely separated from the sub-leading jet. The lepton may come from the decay of a sneutrino or a heavy neutrino, both decays can also contain jets. As the mother particle is boosted if it results from a $Z^{\prime}$ decay, the lepton and the jet(s) tend to have a small angular separation. For the background or a signal event from the SM $Z$ portal such a boost is not typical, so the hardest particles are scattered back-to-back. To understand these distributions better, we also study the angular separation between the lepton and jets in the pseudo-rapidity and azimuthal planes separately, see Figs. \ref{fig:Delta-phi} and \ref{fig:Delta-eta}, respectively. In both figures, we observe that the signal events prefer to lie within $|\Delta \phi|<1.5$ for all three variables, while the background events contribute more to the higher values of $|\Delta \phi|$. Almost similar behavior can be seen for $|\Delta \eta|$, where signal events prefer to stay closer while background events have a somewhat wider distribution. 

%--------------------------------------
%--------------------------------------
%%%%%%%%%%%%%%%%%%%%%%%%%%%%%%%%%%%%%%%%%%%%%%%

%--------------------------------------
%--------------------------------------
%%%%%%%%%%%%%%%%%%%%%%%%%%%%%%%%%%%%%%%%%%%%%%%

%--------------------------------------
%--------------------------------------
%==========================================================
\subsection{T-odd observables}

CPV can manifest itself as an asymmetry that results in an observable difference in angular distributions of the decay products of a particle and its antiparticle. The scalar Triple products (TPs) are designed to probe such asymmetries. Here we construct three TPs, primarily using the 3-momentum of the leading lepton and the two leading jets. Furthermore, we also calculate similar TPs using the two MET components: it is therefore a 2-dimensional projection of the original 3-dimensional TPs.

\begin{figure}[htb!]
\centering
\begin{subfigure}{0.33\textwidth}
  \centering
  \includegraphics[width=1\linewidth]{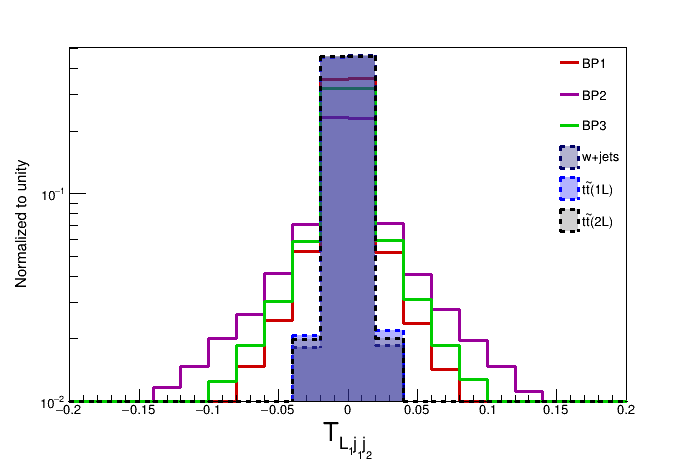}
\end{subfigure}%
\begin{subfigure}{.33\textwidth}
  \centering
  \includegraphics[width=1\linewidth]{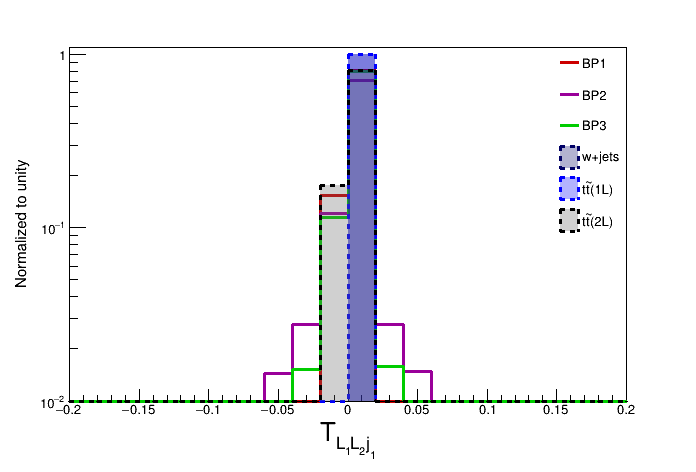}
\end{subfigure}
\begin{subfigure}{.33\textwidth}
  \centering
  \includegraphics[width=1\linewidth]{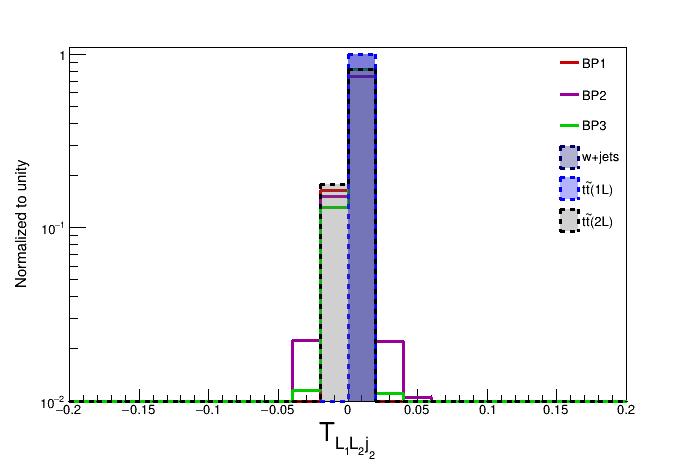}
\end{subfigure}
\caption{Distribution of the three T-odd observables that are calculated without using the MET.}
\label{fig:Toddlep}
\end{figure}

\begin{figure}[htb!]
\centering
\begin{subfigure}{0.33\textwidth}
  \centering
  \includegraphics[width=1\linewidth]{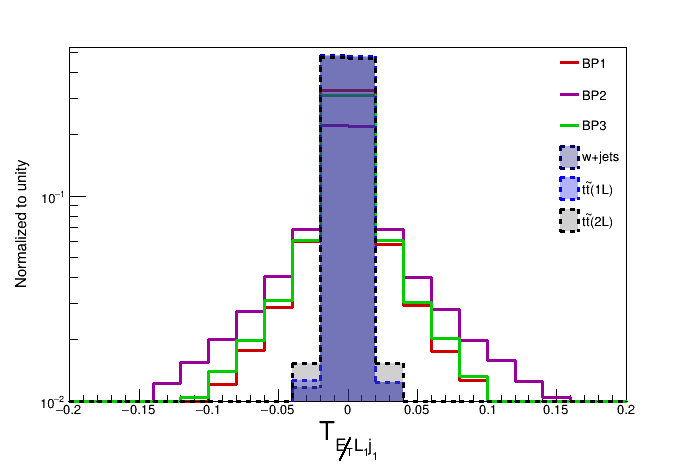}
  %\caption{Obsx1j-CPV VS SM}
  %\label{fig:sub1}
\end{subfigure}%
\begin{subfigure}{.33\textwidth}
  \centering
  \includegraphics[width=1\linewidth]{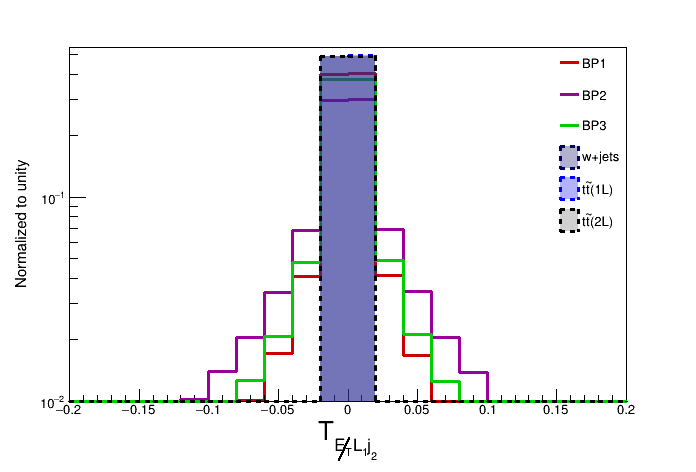}
  %\caption{Obsx2j-CPV VS SM}
  %\label{fig:sub2}
\end{subfigure}
\begin{subfigure}{.33\textwidth}
  \centering
  \includegraphics[width=1\linewidth]{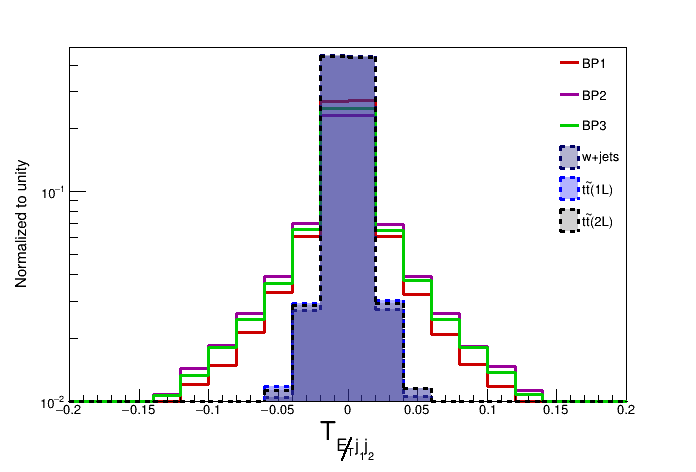}
  %\caption{Obsx3j-CPV VS SM}
  %\label{fig:sub2}
\end{subfigure}
\caption{Distribution of the set of T-odd observables that are calculated using the MET.}
\label{fig:Toddmet}
\end{figure}

The T-odd triple products used in this work are listed below, 
\begin{itemize}
    \item $T_{L_1j_1j_2}= \frac{p_{L_1}\cdot (p_{j_1}\times p_{j_2})}{s^{3/2}}$, \quad $T_{L_1L_2j_1}=\frac{p_{L_1}\cdot(p_{L_2}\times p_{j_1})}{s^{3/2}}$, \quad and \quad $T_{L_1L_2j_2}=\frac{p_{L_1}\cdot(p_{L_2}\times p_{j_2})}{s^{3/2}}$

    \item $T_{\slashed{E}_{T}L_1j_1} =\frac{\slashed{E}_{T}\cdot(p_{L_1}~\times p_{j_1})}{s^{3/2}}$, \quad $T_{\slashed{E}_{T}L_1j_2} =\frac{\slashed{E}_{T}\cdot(p_{L_1}~\times p_{j_2})}{s^{3/2}}$, \quad and \quad $T_{\slashed{E}_{T}j_1j_2} =\frac{\slashed{E}_{T}\cdot(p_{j_1}~\times p_{j_2})}{s^{3/2}}$,
\end{itemize}
where, $p_{L_1}$, $p_{j_1}$, and $p_{j_2}$ denote the 3-momentum of the lepton, leading jet, and sub-leading jet, respectively. To convert these observables into dimensionless quantities, we normalize them with $s^{3/2}$ where $\sqrt{s}=100~\rm TeV$ is the center of mass energy. These quantities are plotted in Figs.  \ref{fig:Toddlep} and \ref{fig:Toddmet}. We may see that the distribution of the signal benchmarks is much wider than that of the SM processes. This is partially due to the harder final-state particles arising from the decays of $Z^{\prime}$ to sneutrinos and partially due to the widening of the distributions caused by CPV. Notice that the distributions in Figs.  \ref{fig:Toddlep} and \ref{fig:Toddmet} are normalised to one, \textit{i.e.} when normalised to luminosity the tails arising from the signal are not as prominent as we shall discuss below. Nevertheless, the fact that the bulk of the SM background has low absolute values of the TP observables gives us some chance of analyzing the signal and the CPV within it.

%---------------------------------------------

%---------------------------------------------
\begin{figure}
    \centering
    \includegraphics[width=0.4\textwidth]{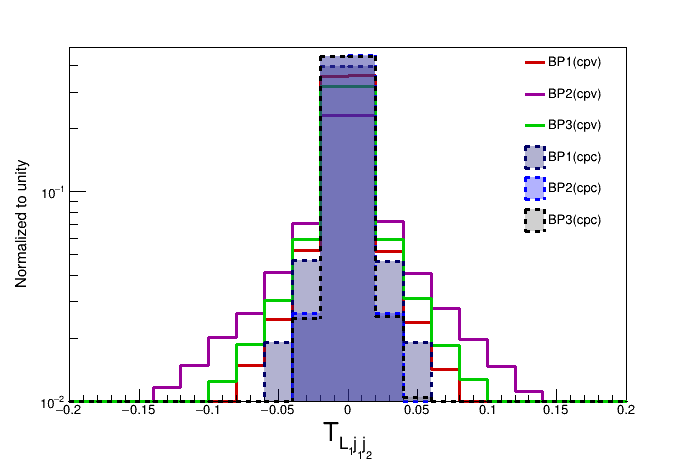}
    \includegraphics[width=0.4\textwidth]{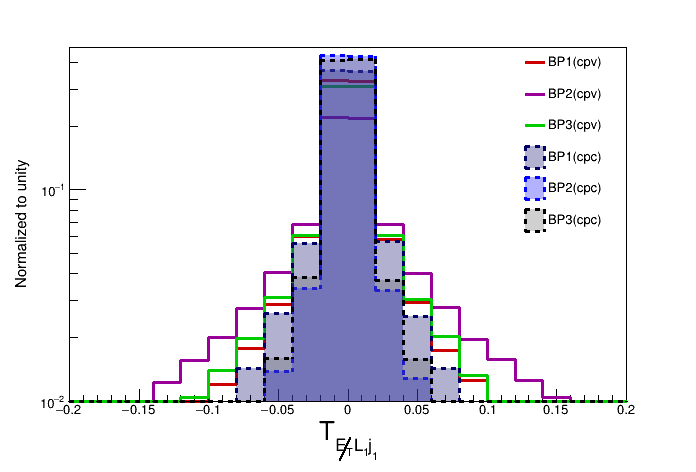}
    \caption{Distributions of T-odd observables for CP violating benchmark points compared to CP conserving ones. We may notice that the distributions are wider with CP violation.}
    \label{fig:CPVvsCPC}
\end{figure}

%\vspace{2cm}
%------------------------------------------
In section \ref{sec:CPV} we conjectured that the width of the distributions of T-odd observables would be wider if CPV effects were present compared to a similar CP conserving case. In Fig. \ref{fig:CPVvsCPC} we show a comparison of the distributions of the T-odd observables for CP conserving and violating benchmarks. This indeed seems to confirm our conjecture: CP violation widens the distributions of T-odd observables, and the difference seems to be substantial. 

To quantify the separation between the SM values and the BSM ones, we calculate the area covered by a given distribution outside a specific range of the T-odd observable. The higher the region outside the window of the T-odd observable, the stronger the impact of the CPV phase on the T-odd observable. In Tab. \ref{tab:area-under-curve}, we show the area of the tails estimated using the following equation: 
\begin{equation}
\mathcal{A_T} = 1 - \frac{\rm Area~with~|T|<0.06}{\rm Total~area~}. 
\label{asymmetry}
\end{equation}
From the data displayed in Tab. \ref{tab:area-under-curve}, one can observe the impact of the CP violation and the harder final-state objects for the signal BPs. The SM background consists of softer objects and the distributions do not get much widened by the CKM phase as its effects get suppressed by the small mixing angles of the CKM matrix.  In the next section, we will study the impact of different kinematic variables on these T-odd observables and make use of them to obtain a good signal over background ratios.  

%-------------------------------------------
\begin{table}[htb!]
    \centering
    \begin{tabular}{|c|c|c|c|c|c|c|} \hline
      \makecell{T-odd \\ observable} &  BP1 & BP2 & BP3 & $W$+jets & $t\Bar{t}$(1L) & $t\Bar{t}$(2L)  \\ \hline
       $T_{L_1j_1j_2}$  & 0.114 & 0.279 & 0.157 & 0.025 & 0.025 & 0.022\\ \hline
       
       $T_{L_1L_2j_1}$  & 0.008 & 0.074 & 0.033 & 0.0 & 0.0 & 0.002\\ \hline
       
       $T_{L_1L_2j_2}$  & 0.002 & 0.037 & 0.016 & 0.0 & 0.0 & $5.8\times 10^{-4}$ \\ \hline
       
       $T_{\slashed{E}_{T}L_1j_1}$  & 0.149 & 0.310 & 0.176 & 0.007 & 0.005 & 0.01  \\ \hline
       $T_{\slashed{E}_{T}L_1j_2}$  & 0.069 & 0.169 & 0.091 & 0.003 & 0.002 & 0.003 \\ \hline
       $T_{\slashed{E}_{T}j_1j_2}$  & 0.241 & 0.286 & 0.267 & 0.038 & 0.038 & 0.039\\ \hline
\end{tabular}
\caption{{ The area of the distribution tails $\mathcal{A_T}$ calculated using the preselection on the given T-odd observable, as defined in Equation (\ref{asymmetry}).}}
\label{tab:area-under-curve}
\end{table}

%%%%%%%%%%%%%%%%%%%%%%%%%%%%%%%%%%%%%%%%%%%%%%%%%%%%%%%%%%%%%%%%

\subsection{Cut-based analysis}

We first have a look at what kind of results can be obtained through a classical cut-based analysis, using standard kinematical variables and the distributions of the T-odd observables introduced above. We shall improve this strategy using boosted decision trees in the following subsection.

As we discussed above, CP violation makes the distributions of the T-odd observables wider than it would be without it. For the SM backgrounds the distribution is centered around zero, even more than for the CP conserving versions of our benchmarks, see Figs. \ref{fig:Toddlep}--\ref{fig:CPVvsCPC}. Here we analyze the prospects for a cut-and-count analysis for BP2 which, following the Tab.  \ref{tab:BPs} and Tab. \ref{tab:BR}, seems to have the largest impact on the additional CP phases. 

\begin{figure}[htb!]
    \centering
    \includegraphics[scale=0.3]{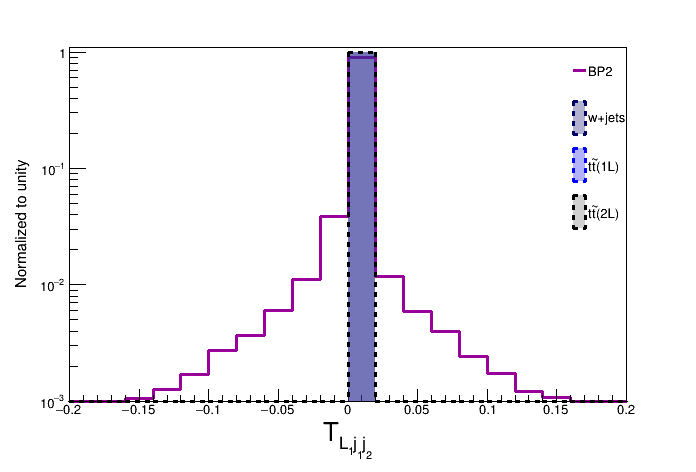}
    \caption{The distribution of the T-odd observable $T_{L_1j_1j_2}$ after the imposition of the kinematic cuts $p_{T}^{j_1}>300$ GeV, $p_{T}^{j_2}>150$ GeV, $\Delta R_{L_1j_2}<1.5$, and $\Delta R_{j_1j_2}<1.5$.}
    \label{fig:Tlj1j2-cut}
\end{figure}

The two T-odd observables that show the best sensitivity are 
$T_{L_1j_1j_2}$ and $T_{\slashed{E}_{T}L_1j_1}$ (see Tab. \ref{tab:area-under-curve}). However, to maximize the impact of these observables, we first impose the cuts on the kinematic variables and then focus on different regions of the T-odd observables.  The observable $T_{L_1j_1j_2}$ depends on the transverse momentum of the leading and sub-leading jets along with leading lepton. Following the differences observed in the $p_T$ distribution of the jets (see the middle and right panel of Fig.  \ref{fig:jetPT}, respectively), we select events with $p_T^{j_1}$ larger than 300 GeV and $p^{j_2}_T > 150$~GeV. We further constrain the angular separation of the leading lepton and the sub-leading jet  (middle panel of Fig. \ref{fig:Delta-r}) by selecting events with $\Delta R_{L_1j_2} \leq 1.5$. Furthermore, we also use a cut on the angular separation between the two jets i.e., $\Delta R_{j_1j_2}<1.5$. After imposing these  cuts, we plot the T-odd observable as shown in Fig. \ref{fig:Tlj1j2-cut}, which now clearly shows the minimal contamination of the SM backgrounds. Using this distribution, we now proceed to calculate the signal significance $\mathcal S = \frac{S}{\sqrt{S+B}}$, where S and B represent the number of signal and background events at the integrated luminosity $\mathcal L = 3 ~{\rm ab^{-1}}$. In Tab. \ref{tab:Tlj1j2-cut}, we show how significance changes with the different choices of cuts on the T-odd observable. We use four specific regions of the observable, namely $|T_{L_1j_1j_2}|>$ 0.001, 0.01, 0.05, and 0.1. As the $|T_{L_1j_2j_1}|$ increases, the number of signal events that pass the cut reduces, and the same for the background events also reduces significantly, and therefore, signal significance increases.

%%%%%%%%%%%%%%%%%%%%%%%%%%%%%%%%%%%%%%%%%%%%%%%%%%%%%%%%%%
%%%New Plot and Table start%%%%%%%%%%%%%%%%%%%%%%%%%%%%%%%%%%%%
\begin{table}[htb!]
    \centering
    \begin{tabular}{|c|c|c|c|c|c|c|} \hline
       $|T_{L_1j_1j_2}|$  & Signal & $W$+jets & $t\bar{t}$(1L) &  $t\bar{t}$(2L) & Total bgnd & $\frac{S}{\sqrt{S+B}}$ \\ \hline
        $>0.001$  & 9220  & $25.3\times 10^7$  & $27.1\times 10^7$ & $2.56\times 10^7$ & $54.96\times 10^7$ & 0.39 \\ \hline
        $>0.01$  & 7230 & $15.8\times 10^7$ & $16.5\times 10^7$  & $1.64\times 10^7$ & $33.94\times 10^7$ & 0.39 \\ \hline
        $>0.05$  & 4150 & $5.15\times 10^7$ & $5.81\times 10^7$  & $0.62\times 10^7$ & $10.10\times 10^7$ & 0.41\\ \hline
        $>0.10$  & 2860 & $1.19\times 10^7$ & $2.03\times 10^7$  & $0.32\times 10^7$ & $3.54\times 10^7$ & 0.48\\ \hline
    \end{tabular}
    \caption{{The signal significances after the cuts as well as a cut on the T-odd observable $T_{L_1j_1j_2}$. For the signal events we use BP2. The integrated luminosity is 3 ab$^{-1}$ and the collision energy $\sqrt{s}=$100 TeV. }}
    \label{tab:Tlj1j2-cut}
\end{table}

\begin{figure}[htb!]
    \centering
    \includegraphics[scale=0.3]{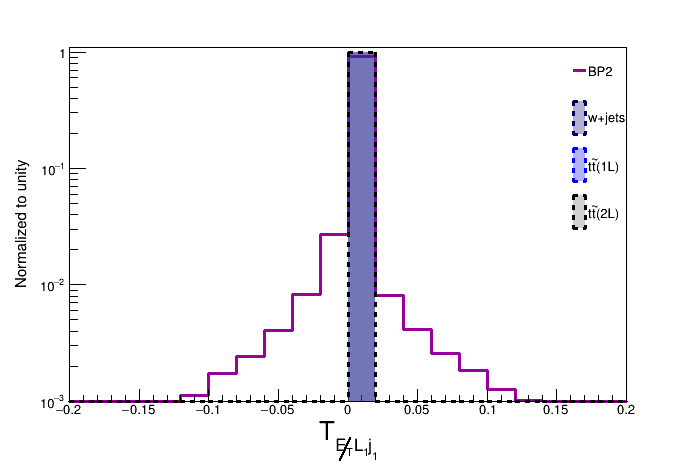}
    \caption{{The distribution of the T-odd observable $T_{\slashed{E}_TL_1j_1}$ after the application of the kinematic cuts $\slashed{E}_{T}>300~{\rm GeV}$, $p^{j_1}_{T}>300~{\rm GeV}$ and $\Delta R_{L_1j_1}<1.5$.}}
    \label{fig:Tplj1-cut}
\end{figure}

The second best T-odd observable, based on Tab. \ref{tab:area-under-curve}, is $T_{\slashed{E}_TL_1j_1}$, which involves the leading jet, leading lepton and  MET. Following the strategy adopted for $T_{L_1j_1j_2}$, we first select events based on the kinematic cuts and later analyze the cuts on $T_{\slashed{E}_TL_1j_1}$. The cuts placed on the relevant variables are as follows: $\slashed{E}_T>300~\rm GeV$, $ p^{j_1}_T > 300$~GeV and $\Delta R_{L_1j_1} \leq 1.5$. The distribution of $T_{\slashed{E}_TL_1j_1}$ seems prominent after the imposition of the above-mentioned cuts, as shown in Fig. \ref{fig:Tplj1-cut}. The calculation of signal significances with the different binned selection of $T_{\slashed{E}_TL_1j_1}$ is shown in Tab. \ref{tab:Tplj1-cut}. 

We may note that while the distribution of the T-odd observables for the signal is wider, the background has a cross section that is several orders of magnitude larger. Even though the SM background is strongly centered around zero, there is a minor tail that is larger than that coming from the signal. Hence the significance of the deviation at the tails remains at a modest level. 

A cut-based selection can be considered to be a decision tree. The efficiency of such a selection can be enhanced by using machine learning to decide the optimal boundary. Furthermore, boosting uses an ensemble of shallow decision trees trained by reweighting incorrectly identified data points and takes the average of the decision boundaries of such trees \cite{smlbook}. This allows for more complicated decision boundaries and usually leads to a better discrimination power.
Therefore, we perform a Multi-Variate Analysis (MVA), namely the BDT method implemented in 
``Toolkit for Multivariate Analysis" (TMVA) framework \cite{TMVA:2007ngy}, as discussed in the next section. Unlike cut-based approaches, a BDT can effectively capture complex correlations and nonlinear relationships within the data, allowing for more nuanced discrimination between signal and background events, and thereby elevating the overall significance.

%--------------------------------------------------

\begin{table}[htb!]
\centering
\begin{tabular}{|c|c|c|c|c|c|c|} \hline
 $|T_{\slashed{E}_{T}L_1j_1}|$   & Signal  & $W$+jets & $t\bar{t}$(1L) & $t\bar{t}$(2L) & Total bgnd & $\frac{S}{\sqrt{S+B}}$ \\ \hline
$>0.001$ & 9130 & $9.68\times 10^7$ & $5.36\times 10^7$ & $0.85\times 10^7$ & $15.89\times 10^7$ & 0.72 \\ \hline
$>0.01$  & 7030 & $5.28\times 10^7$ & $3.11\times 10^7$ & $0.47\times 10^7$ & $8.86\times 10^7$ & 0.74 \\ \hline
$>0.05$  & 4040 & $0.88\times 10^7$ & $0.87\times 10^7$ & $0.14\times 10^7$ & $1.89\times 10^7$ &  0.93 \\ \hline
$>0.10$  & 2640 & $0.30\times 10^7$ & $0.25\times 10^7$ & $0.07\times 10^7$ & $0.65\times 10^7$ & 1.04 \\ \hline
\end{tabular}
\caption{{The signal significances after the cuts as well as the cut on the T-odd observable $T_{\slashed{E}_{T}L_1j_1}$. For the signal events we use BP2 as in Tab. \ref{tab:Tlj1j2-cut}. }}
\label{tab:Tplj1-cut}
\end{table}

%=========================================================================

%\end{itemize}

\subsection{BDT analysis}

The cut-based method focuses on the identification of the zones in the feature space by drawing rectangular boundaries on the features to increase the signal efficiency over the background. This approach has its limitations. The use of a BDT trained with samples of (simulated) signal and background data results in a more complex decision boundary between the classes and hence generally improves the performance in classifying the data. To bypass the shortcomings of the simple cut-based analysis, we use a BDT where we consider all the features or observables in a single go and obtain a Receiver Operating Characteristics (ROC) curve giving the probability of correctly rejecting background events as a function of the probability of correctly identifying signal events. We use the ROOT Library, ``Toolkit for Multivariate Analysis" \cite{TMVA:2007ngy} to perform the MVA. More specifically, we use the "Adaptive Boosted Decision Tree" (AdaBoost) algorithm \cite{FREUND1997119} for the classification of signal and background events\footnote{
Different SM backgrounds are weighted according to their cross sections while performing the BDT analysis.}. The AdaBoost classifier algorithm constructs various dependency matrices \cite{book} based on the available features of the dataset provided and arranges these features according to their respective importance for classification. 

\begin{figure}[htb!]
    \centering
    \includegraphics[scale=0.5]{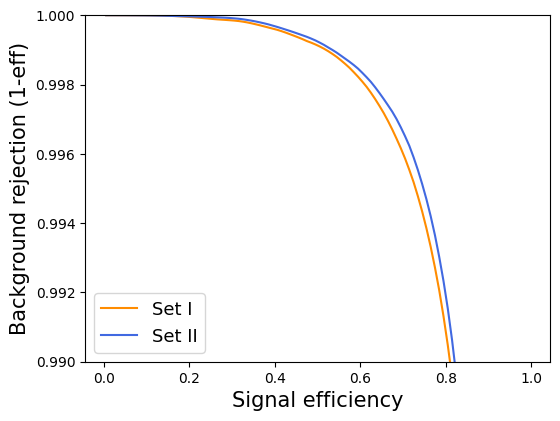}
    \caption{{ The receiver operating characteristic (ROC) curve obtained using the AdaBoost algorithm. The two above-mentioned sets were used for the input of BDT algorithm for classifying signal events (BP2) from the SM processes. }}
    \label{fig:roc-set1-set2}
\end{figure}

Two sets of inputs are provided in the dataset: Set-I includes all the relevant kinematic variables, while Set-II includes all the T-odd observables in addition to the kinematic variables of Set-I. \ref{tab:set}, we list all these variables in Set I and Set II. In Fig. \ref{fig:roc-set1-set2}, we show the ROC curve for the two sets of input corresponding to BP2. Any difference in the ROC curves for the two sets is an improvement of the classification due to the T-odd observables. The blue line represents the ROC curve for Set II, which clearly shows the impact of the T-odd observables in improving the signal efficiencies. The elbow point corresponds to the coordinates of 60\% signal efficiency and a mistagging rate of 0.1\%.

\begin{table}[htb!]
    \centering
    \begin{tabular}{|c|c|} \hline
        Set I & $\slashed{E}_T$, $M_{\rm eff}$, $\slashed{E}_T/\sqrt{H_T}$, $H_T$, $M_{T}^{L\nu}$, $M_{L_1j_1j_2}$ ,$\Delta R_{lj_1}$, $\Delta R_{lj_2}$, $\Delta R_{j_1j_2}$, $p_T^L$ ,$p^{j_1}_T$, $p^{j_2}_T$  \\ [3mm] 
        \hline 
        Set II & \makecell{$\slashed{E}_T$, $M_{\rm eff}$, $\slashed{E}_T/\sqrt{H_T}$, $H_T$, $M_{T}^{L\nu}$, $M_{L_1j_1j_2}$ ,$\Delta R_{lj_1}$, $\Delta R_{lj_2}$, $\Delta R_{j_1j_2}$, $p_T^L$ ,$p^{j_1}_T$, $p^{j_2}_T$ \\ $T_{\slashed{E}_{T}L_1j_1}$, $T_{\slashed{E}_{T}L_2j_2}$, $T_{\slashed{E}_{T}j_1j_2}$, $T_{L_1j_1j_2}$, $T_{L_1L_2j_1}$, $T_{L_1L_2j_2}$}\\ [4mm]
        \hline 
    \end{tabular}
    \caption{The two sets of input variables used in the BDT analysis.}
    \label{tab:set}
\end{table}

\begin{figure}[htb!]
\centering
  \includegraphics[scale=0.5]{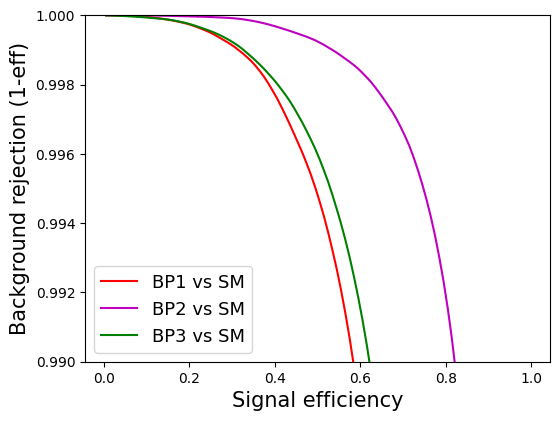}
  \includegraphics[scale=0.48]{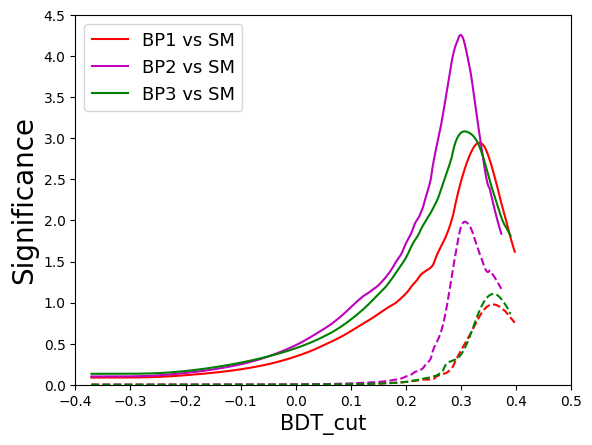}
\caption{{Left: The receiver operating characteristic (ROC) curve obtained using Set II as BDT inputs for BP1, BP2 and BP3. Right: The signal significance with the variation of BDT-cut for the three BPs. The dashed lines show the significance measured by the Asimov significance Eq. \ref{eq:Asimov sig} with the background uncertainty $\sigma_b/B = 1 \%$. While the solid lines describe the significance with zero uncertainty. In both cases, the integrated luminosity $\mathcal{L}=3~ \rm ab^{-1}$ is considered.}}
\label{fig:significance}
\end{figure}

To better represent the effect of the T-odd observables, we perform the BDT analysis for all BPs using the input variables of Set-II. The performance of the classifier is shown in Fig. \ref{fig:significance} (left). Among the three BPs related to new CPV phases, the BP2 captures the maximum CPV effect leading to wider distributions of the T-odd observables. This results in a higher signal efficiency for BP2 as shown by the solid magenta line in Fig. \ref{fig:significance} (left). To quantify the impact of these new variables on the prospect of discovery/exclusion of this new physics scenario, we use the Asimov estimate of statistical significance. An important feature of this kind of significance measure is that it allows the inclusion of systematic uncertainties; for details, we refer to \cite{Cowan:2010js, Cowan:2017}. For the case of Poisson distributed signal ($S$) and background ($B$) events with background uncertainty $\sigma_b$, the approximated statistical significance is 
\begin{equation}
    \mathcal{S}=\sqrt{2\Bigg((S+B)~ln\Big[\frac{(S+B)(B+\sigma_b^2)}{B^2+(S+B)\sigma_b^2}\Big]-\frac{B^2}{\sigma_b^2}~ln\Big[1+\frac{\sigma_b^2 S}{B(B+\sigma_b^2)}\Big]\Bigg)}. \label{eq:Asimov sig}
\end{equation}
In the case where the background is known exactly ($\sigma_b$ = 0) and/or $\sigma_b << B$, then this simplifies to
\begin{equation}
    \mathcal{S} = \frac{S}{\sqrt{S+B}}. \label{eq:sig}
\end{equation}
The systematic uncertainty $\sigma_b$ is usually assumed to be proportional to $\kappa$-th fraction of the total background events ($B$), where $\kappa$ is a free parameter ranging over the interval  [0,1]. 

\begin{figure}[h!]
    \centering
    \includegraphics[width=0.5\linewidth]{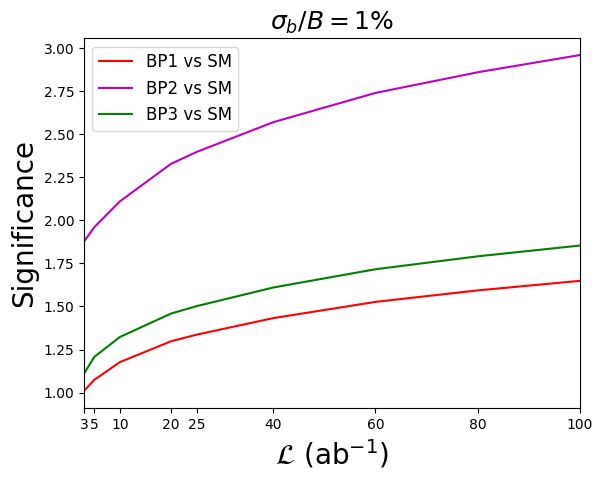}
    \caption{The significance varies with the integrated luminosity $\mathcal{L}$. Here the systematic uncertainty $\sigma_b/B = 1\%$ is considered into the Asimov significance Eq. \ref{eq:Asimov sig}.}
    \label{fig:sig-vs-lumi}
\end{figure}

We calculate the signal significance for all BPs by varying the cut on the BDT score, which essentially means calculating the signal efficiencies at different values of mis-tagging rates. The right panel of Fig.  \ref{fig:significance} displays the variation of signal significance with the BDT cut values. As evident in the figure, we can achieve around 4$\sigma$ statistical significance at the integrated luminosity $\mathcal{L}=3~ \rm ab^{-1}$ (with $\sigma_b = 0$) for BP2 while the other two BPs give lower significances. However, if we assume the systematic uncertainty to be  $1\%$, i.e., $\frac{\sigma_b}{B} = 1\%$, the significance is reduced to around 2$\sigma$, as shown by the magenta dashed line in the same graph. A similar reduction in significance is observed for the other two BPs as well. Furthermore, the sensitivity degrades further if we include more systematic uncertainty in our analysis, as is evident from Tab. \ref{tab:signi-uncertainty}. A few points worth mentioning here are as follows. First, note that the (expected) maximum reach of the FCC-hh is $100 ~\rm ab^{-1}$, as mentioned in \cite{Zimmermann:2016puu}. Secondly, given the ongoing paradigm shift with the advent of Machine Learning (ML) algorithms, significant improvement in our understanding of different collider objects and overall technical improvements are envisioned, therefore, we expect that much more data will be collected at the end of the FCC-hh, so that the prospect of discovering this scenario by that time will be increased, as tentatively shown in Fig. \ref{fig:sig-vs-lumi}. 

%
%Thus the right panel of Fig. \ref{fig:significance} shows the comparison of the significance for different benchmark points and depicts the significance's dependency on the background uncertainty $\sigma_b$. The two equations Eq. \ref{eq:Asimov sig} (with the information of the uncertainty measurements) and Eq. \ref{eq:sig} (without the uncertainty) are shown in this Fig. \ref{fig:significance} (right panel) by the dashed and solid lines respectively. 

%Furthermore, we also checked the significance, with the variation of the integrated luminosity. As mentioned in \cite{Zimmermann:2016puu}, the maximum reach of the future high-energy frontier circular collider is $100 ~\rm ab^{-1}$, we show the variation of significance with the luminosity up to $100 ~\rm ab^{-1}$ in Fig. \ref{fig:sig-vs-lumi}. Fig. \ref{fig:sig-vs-lumi} shows that to achieve a higher significance the luminosity is expected to be higher. \\

%\newpage
%In Table \ref{tab:signi-uncertainty}, we compare the BDT\_cut values for the maximum significance and the maximum significance values for different choices of the systematic background uncertainty for all the BPs. Here we followed Eq. \ref{eq:Asimov sig} for particular choice of luminosity $\mathcal{L}= 3 ~\rm ab^{-1}$.

\begin{table}[htb!]
    \centering
    \begin{tabular}{|c|c|c|c|} \hline
        $\sigma_b/B$ & BP & BDT\_cut & Significance \\ \hline
         & $\rm BP1$ & 0.330 & 2.9 \\ \cline{2-4} 
       $0\%$ & $\rm BP2$ & 0.280 & 4.4 \\ \cline{2-4} 
        & $\rm BP3$ & 0.304 & 3.1\\ \hline \hline
         & $\rm BP1$ & 0.331 & 1.0 \\ \cline{2-4} 
       $1\%$ & $\rm BP2$ & 0.284 & 1.9\\ \cline{2-4} 
        & $\rm BP3$ & 0.329 & 1.1\\ \hline  \hline  
         & $\rm BP1$ & 0.331 &  0.2\\ \cline{2-4} 
       $10\%$ & $\rm BP2$ & 0.284 & 0.3 \\ \cline{2-4} 
        & $\rm BP3$ & 0.329 & 0.2\\ \hline        
    \end{tabular}
    \caption{Comparison of the statistical significance for different choices of the systematic uncertainty at integrated luminosity $\mathcal{L}= 3 ~\rm ab^{-1}$. }
    \label{tab:signi-uncertainty}
\end{table}

In using machine learning methods, we have to use simulated samples of data both for the SM and our BSM signal to train the BDT. In our case, we of course know the actual features of our benchmark points and can generate accurate BSM data. One may ask how well would this be justified if we were to investigate real data?

In investigating CPV we are doing a characterization of BSM physics. This means that we should have discovered it first. In the BLSSM with $Z^{\prime}$ as a portal to sneutrino production, we would probably have seen $Z^{\prime}$ as a dilepton and dijet resonance, which would give us its mass. Similarly, it seems reasonable that we would have an idea of the spectrum of the superpartners, which determines the decay kinematics. Obviously, there would be some remaining uncertainty in these, which will reduce the discriminating power of the BDT. However, we expect the uncertainties in the BSM spectrum to be so small that the reduction in significance would be small.

%\newpage 
%----------------------------------------------------
\subsection{The CPC limit and $Z^{\prime}$ impact}\label{sec:Zprime}

\begin{figure}[htb!]
    \centering
    \includegraphics[scale=0.5]{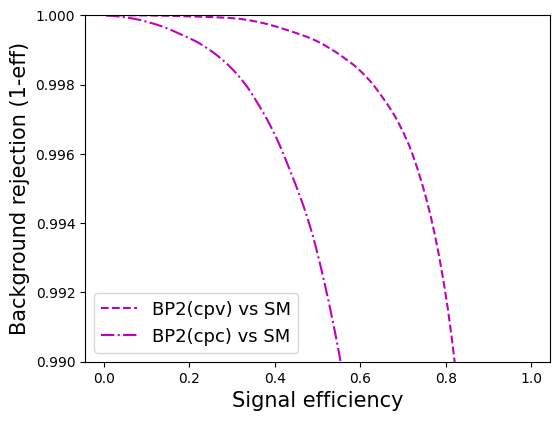}
    \includegraphics[scale=0.5]{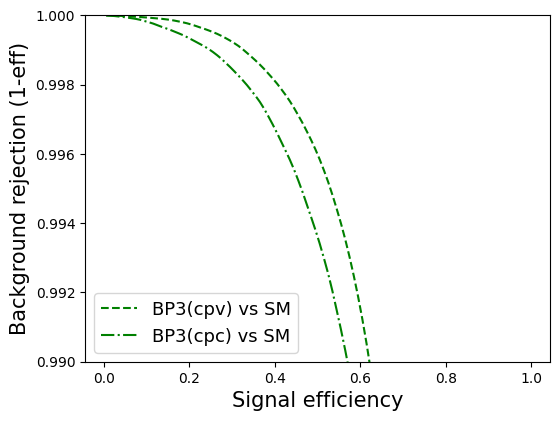}
    \caption{Comparison of the ROC curves for BP2 (left panel) and BP3 (right panel) with CPV ({dashed} line) and CPC ({dot-dashed} line) cases. In both cases, all the SM processes are used as the background for the BDT analysis.}
    \label{fig:roc-cpc-cpv}
\end{figure}

%===============================
Until now, we have studied the impact of the new CPV phases through the T-odd observables. In the previous subsection, we showed that we can distinguish the signal from the SM. We should also see that we can establish that there is CPV by comparing the CPV case to a corresponding CPC benchmark. We perform the previously discussed BDT analysis using all the observables presented as Set-II, our findings are shown in Fig. \ref{fig:roc-cpc-cpv}. Here we present a comparative analysis of the ROC curves for the CPC and CPV scenarios for the selected reference points, BP2 and BP3. As indicated in the Tab. \ref{tab:area-under-curve}, these two benchmark points exhibit greater asymmetry compared to BP1. The magenta and green {dot-dashed} lines correspond to the cases when BP2 and BP3 are considered as the signal setting all new phases to zero (keeping the CKM phase to the desired non-zero value), while the { dashed} lines denote the same BP but with non-zero CPV phases. Comparing the ROC curves of the CPV and CPC cases, we can see that it is much easier to distinguish a CPV benchmark from the SM than a CPC one. 

%=============================

\begin{figure}[htb!]
    \centering
    \includegraphics[width=0.45\textwidth]{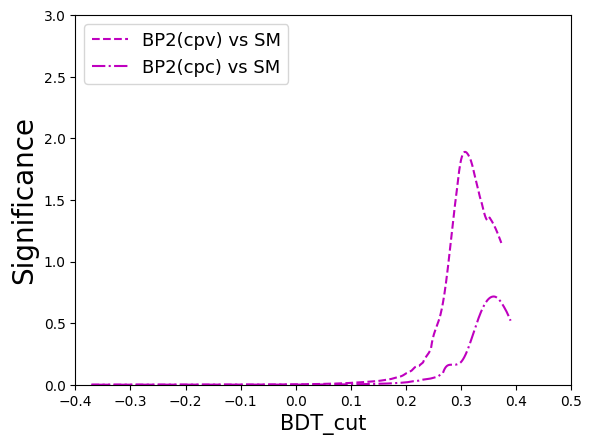}
    \includegraphics[width=0.45\textwidth]{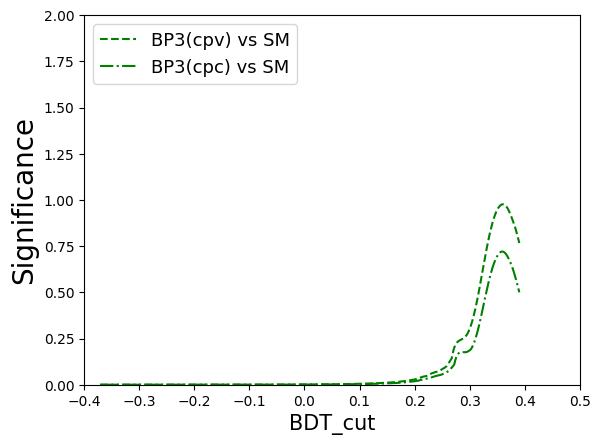}
    \caption{Comparison of the statistical significance curves for BP2 (left panel) and BP3 (right panel) with CPV ({dashed} line) and CPC ({dot-dashed} line) cases. The Asimov significance Eq. \ref{eq:Asimov sig} with systematic background uncertainty $\sigma_b/B = 1\%$ is considered along with integrated luminosity $\mathcal{L} = 3 ~\rm ab^{-1}$.}
    \label{fig:sig-cpc-cpv}
\end{figure}

To quantify the statement made above while comparing the ROC curves for the CPV and CPC cases as in Fig. \ref{fig:roc-cpc-cpv}, we display the statistical significance curves for BP2 and BP3 in Fig. \ref{fig:sig-cpc-cpv}. The CPC scenario exhibits significantly lower significance compared to the CPV scenario. Specifically, the significance curves for the CPC scenario, for both BP2 and BP3, are around $0.5 \sigma$. In contrast, the significance for the CPV scenario is around $2\sigma$ for BP2 and $1\sigma$ for BP3. The difference between the significances between the CPC and CPV cases for BP2 is greater than $1\sigma$, so in the case of large CPV, we can at least reject a CP-conserving scenario with more than $95\%$ confidence. For BP3 with a smaller CPV one cannot make such a claim.

Another point of worth mentioning is the role played by the mediator particle $Z^\prime$ in our analysis. The pair production of sneutrinos through proton-proton collisions involves both the SM Z boson and the $Z^\prime$ boson, a new particle associated with the $B-L$ symmetry. However, the mass of $Z^\prime$ has to be quite large in order to satisfy the existing limits from the collider experiments. Therefore, the higher the mass of $Z^\prime$ (around 5 TeV), lower the contribution to the total cross section. However, after being produced on-shell at higher energies, it can provide a significant boost to the sneutrinos of mass around a TeV, and thereby its decay products. The T-odd observables are constructed using the 4-momentum of the decay products of sneutrinos, and therefore the presence (or absence) of $Z^{\prime}$ will definitely play a role in the overall outcome. We find that the inclusion of $Z^{\prime}$ in the production process increases the cross section (LO in $\alpha_s$) by a factor of 5 at the 100 TeV collider, as shown in Tab. \ref{tab:BP Zp}.       

\begin{table}[htb!]
    \centering
    \begin{tabular}{|c|c|}
    \hline
        Signal cross section $\sigma_s$ for CPV2 &  0.0064 pb \\ \hline
        Signal cross section $\sigma_s$ for CPV2 without $Z^{\prime}$ &  0.0011 pb \\ \hline    
    \end{tabular}
    \caption{The sneutrino pair production cross section with and without the $Z^{\prime}$ at the 100 TeV collider. }
    \label{tab:BP Zp}
\end{table}

\begin{figure}[htb!]
\centering
\begin{subfigure}{0.33\textwidth}
  \centering
  \includegraphics[width=1\linewidth]{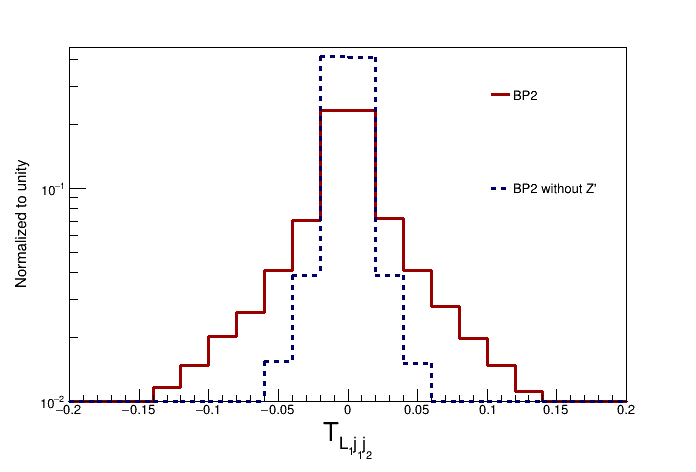}
  %\caption{Obs1j-CPV VS SM}
  %\label{fig:sub1}
\end{subfigure}%
\begin{subfigure}{.33\textwidth}
  \centering
  \includegraphics[width=1\linewidth]{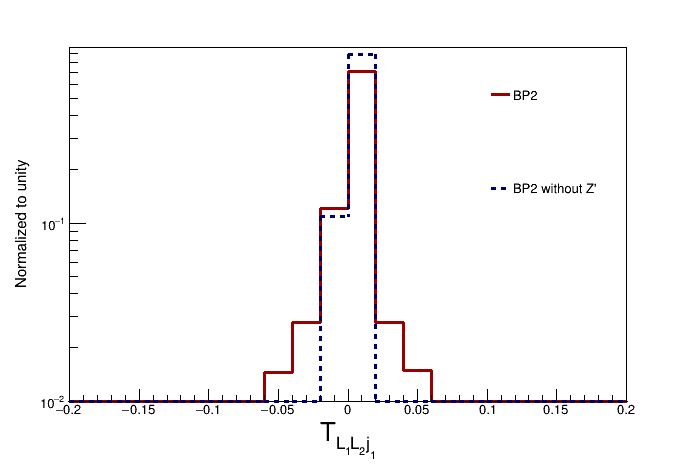}
  %\caption{Obs2j-CPV VS SM}
  %\label{fig:sub2}
\end{subfigure}
\begin{subfigure}{.33\textwidth}
  \centering
  \includegraphics[width=1\linewidth]{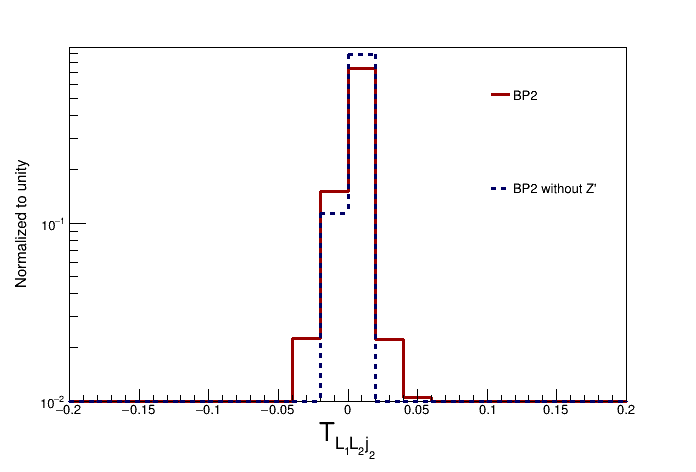}
  %\caption{Obs3j-CPV VS SM}
  %\label{fig:sub2}
\end{subfigure}
\caption{Distribution of the first set of T-odd observables calculated without using MET.}
\label{fig:Obs1-Zp}
\end{figure}

\begin{figure}[htb!]
\centering
\begin{subfigure}{0.33\textwidth}
  \centering
  \includegraphics[width=1\linewidth]{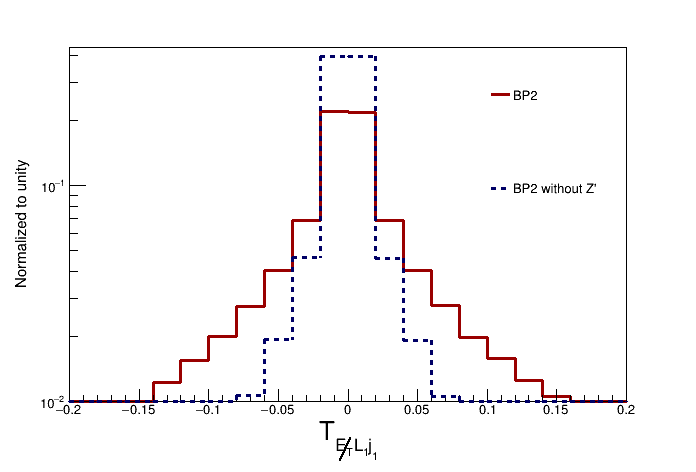}
  %\caption{Obsx1j-CPV VS SM}
  %\label{fig:sub1}
\end{subfigure}%
\begin{subfigure}{.33\textwidth}
  \centering
  \includegraphics[width=1\linewidth]{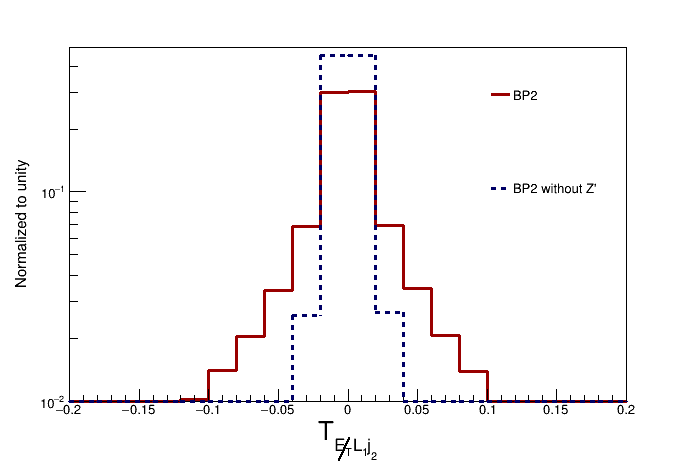}
  %\caption{Obsx2j-CPV VS SM}
  %\label{fig:sub2}
\end{subfigure}
\begin{subfigure}{.33\textwidth}
  \centering
  \includegraphics[width=1\linewidth]{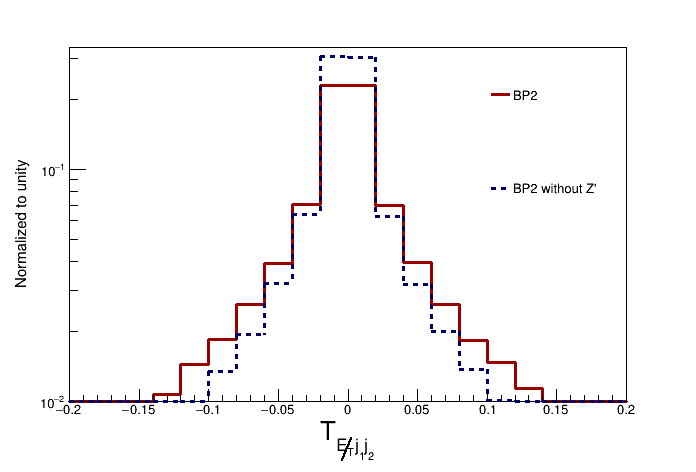}
  %\caption{Obsx3j-CPV VS SM}
  %\label{fig:sub2}
\end{subfigure}
\caption{Distribution of the second set of T-odd observables calculated using MET.}
\label{fig:Obs2-Zp}
\end{figure}

To quantitatively analyze the overall significance of the $Z^\prime$ in the process, a comprehensive analysis of T-odd observables is pursued, elucidating the significance curve to formulate definitive conclusions. As already mentioned, for those signal events without the $Z^\prime$ contribution, the T-odd overvables will be much more restrained around zero, contrary to the case when $Z^\prime$ is present (see Figs. \ref{fig:Obs1-Zp} and \ref{fig:Obs2-Zp} for a visual illustration). 

\begin{figure}[htb!]
\centering
\begin{subfigure}{0.5\textwidth}
  \centering
  \includegraphics[width=0.88\linewidth]{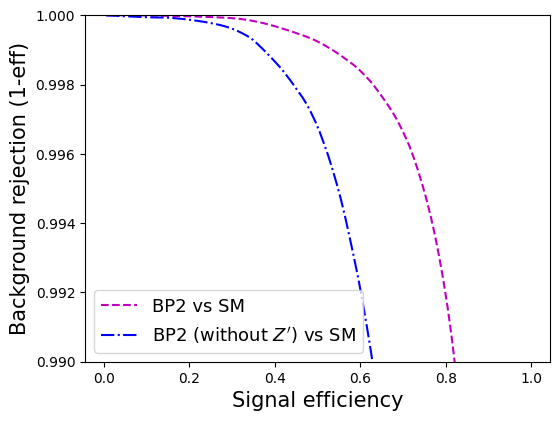}
%\caption{}
%\label{fig:roc-Zp}
\end{subfigure}%
\begin{subfigure}{.5\textwidth}
%  \centering
  \includegraphics[width=0.9\linewidth]{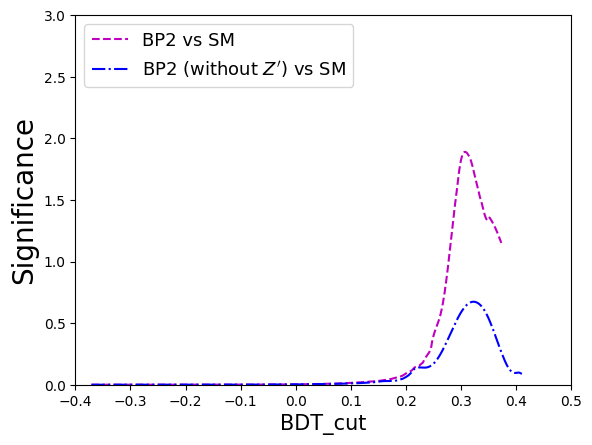}
%\caption{}
%  \label{fig:significance-Zp}
\end{subfigure}
\caption{Left: The ROC curve obtained using the AdaBoost algorithm when the above-mentioned Set II is used for the input for the BP2 (with and without $Z^{\prime}$ separately) and combined SM processes as background. Right: The variation of Signal Significance with the variation of BDT-cut. The Asimov significance Eq. \ref{eq:Asimov sig} with systematic background uncertainty $\sigma_b/B = 1\%$ is considered along with integrated luminosity $\mathcal{L} = 3 ~\rm ab^{-1}$.}
\label{fig:BDT-Zp}
\end{figure}

As discussed earlier, the efficacy of the BDT analysis in discerning signal efficiency over background rejection relies upon the features offered to the BDT. In a similar fashion, we provide the BDT with the same 18 features for the ($p p \rightarrow Z \rightarrow \tilde{\nu} \tilde{\nu}$) process (the case of without $Z^\prime$) along with the previously discussed production process ($p p \rightarrow Z^\prime/Z \rightarrow \tilde{\nu} \tilde{\nu}$) for BP2. Despite the CPV effects originating from the Yukawa matrix elements being the same, the only significant difference is due to the absence of the $Z^\prime$ in the former process. Evidently, the {blue dot-dashed} line of Fig. \ref{fig:BDT-Zp} (left) that shows the signal efficiency over the background rejection rate is reduced if the signal events do not contain the $Z^\prime$ but include only the SM $Z$ boson. At the elbow point, the signal efficiency is approximately 40\% for the process lacking $Z^\prime$, contrasting with the approximately 60\% efficiency observed when $Z^\prime$ is present in the process. Defining $S$ and $B$ as the effective number of signal and background events, respectively, we show the signal significance following the Eq. \ref{eq:Asimov sig} with the variation of the BDT cut in Fig. \ref{fig:BDT-Zp} (right). Due to the diminished cross section, the significance decreases severely for the process without $Z^\prime$. Hence, it is necessary to have an additional gauge portal in the form of the $Z^{\prime}$ boson that is not too heavy to characterize the CPV in the sneutrino sector at the FCC-hh.

{
It is interesting to note that in our work, the $Z^\prime$ boson is produced from the hadron collision of the Future Circular Collider (FCC-hh), leveraging its high energy and wide discovery potential. However, we expect that lepton colliders, viz. International Linear Collider (ILC), Compact Linear Collider (CLIC), Future Circular Collider (FCC-ee), can also play an important and complementary role in the search for $Z^\prime$ bosons, particularly through their capabilities in precise measurements of the relevant couplings. For example, the $Z^\prime$ bosons up to the multi-TeV regime can be effectively probed at the future runs of electron-positron colliders with various center-of-mass energy and using different asymmetry observables; see, for example, \cite{Das:2021esm, Yin:2021rlr}. Furthermore, the proposed Muon collider also has the capability to probe the $Z^\prime$ boson through precision measurements, particularly for masses up to the collider's center-of-mass energy, as noted in \cite{Martinez-Martinez:2023qjt}. If a $Z^{\prime}$ boson was discovered, a dedicated run at the $Z^{\prime}$ pole could allow to study these effects with a better signal-to-background ratio, but lower integrated luminosity compared to future hadron colliders. Also in such an environment a precise prediction of the CPC signal would be crucial to distinguish between CPC and CPV scenarios.}

%%%%%%%%%%%%%%%%%%%%%

\section{Conclusions}

We investigated the possibility of finding CPV in the sneutrino sector at the FCC-hh. We studied this in the BLSSM with an inverse seesaw. This is partially because we needed an efficient portal to produce sneutrinos, which come in the form of the new gauge boson $Z^{\prime}$ related to the U$(1)_{B-L}$ symmetry. The inverse seesaw, on the other hand, allows large superpotential couplings in the (s)neutrino sector, which can be complex and lead to CPV effects. These include direct CPV through the couplings and indirect CPV through the mixing of CP-even and CP-odd sneutrino states.

In general, the effects of CPV can be most clearly seen as a non-vanishing expectation value of T-odd triple product asymmetries. These are products of three three-momenta; in our case a sneutrino could decay to a lepton and two jets. Such an approach would therefore require the ability to distinguish quark jets from antiquark jets, which we did not try to do. Instead, we assumed --- and this assumption was backed by our simulated data --- that there is an intrinsic variance in the distribution of the triple product, which is then widened by the CPV effects. Hence, we could use this widening as an indicator of CPV.

The limits of the BSM searches from the LHC experiments restrict the cross sections for processes involving new physics to be small at future colliders, too. Even though the distributions of our triple-product observables were much wider than for any SM process, the tails of the SM distribution still overshadowed the BSM contribution using a traditional cut-based approach. The significance was too low to claim any kind of evidence for CPV.

The cut-based approach can be improved by using BDTs and this allowed us to distinguish our BSM scenario from the SM background. At least in optimistic cases (large CPV) we could also distinguish between a CPV scenario and a CPC BSM scenario being able to exclude a CP conserving scenario with close to $95\%$ C.L. assuming 1\% systematic uncertainty in the background estimation. This, of course, highlights the need for a sufficiently precise description of the BSM physics in the CPC limit so that the prediction of the distributions of T-odd observables is robust. Given the $2\sigma$ excess already obtained, a somewhat higher luminosity than the one referred to as the baseline would be needed to achieve the desired statistical significance of $3\sigma$.

\section*{Acknowledgements}
SM is financed in part through the NExT Institute and  STFC Consolidated Grant No.
ST/L000296/ 1. HW is supported by the Carl Trygger Foundation under grant No. CTS18:164 and Ruth and Nils-Erik Stenb\"ack's Foundation. Some of the authors acknowledge the use of the IRIDIS High Performance Computing Facility and associated support services at the University of Southampton in the completion of this work. 
The work of AB and AC is funded by the Department of
Science and Technology, Government of India, under Grant No. IFA18-PH 224 (INSPIRE Faculty Award).

%%%%%%%%%%%%%%%%%%%%%%%%%%%%%%%
%\bibliographystyle{unsrt}
%\bibliography{reference}

\begin{thebibliography}{10}

\bibitem{Khalil:2012gs}
S.~Khalil and S.~Moretti.
\newblock {Heavy neutrinos, Z' and Higgs bosons at the LHC: new particles from an old symmetry}.
\newblock {\em J. Mod. Phys.}, 4(1):7--10, 2013.

\bibitem{Khalil:2013in}
S.~Khalil and S.~Moretti.
\newblock {A simple symmetry as a guide toward new physics beyond the Standard Model}.
\newblock {\em Front. in Phys.}, 1:10, 2013.

\bibitem{Wetterich:1981bx}
C.~Wetterich.
\newblock {Neutrino Masses and the Scale of B-L Violation}.
\newblock {\em Nucl. Phys. B}, 187:343--375, 1981.

\bibitem{Buchmuller:1991ce}
W.~Buchmuller, C.~Greub, and P.~Minkowski.
\newblock {Neutrino masses, neutral vector bosons and the scale of B-L breaking}.
\newblock {\em Phys. Lett. B}, 267:395--399, 1991.

\bibitem{Khalil:2007dr}
S.~Khalil and A.~Masiero.
\newblock {Radiative B-L symmetry breaking in supersymmetric models}.
\newblock {\em Phys. Lett. B}, 665:374--377, 2008.

\bibitem{Basso:2008iv}
Lorenzo Basso, Alexander Belyaev, Stefano Moretti, and Claire~H. Shepherd-Themistocleous.
\newblock {Phenomenology of the minimal B-L extension of the Standard model: Z' and neutrinos}.
\newblock {\em Phys. Rev. D}, 80:055030, 2009.

\bibitem{Basso:2009gg}
L.~Basso, A.~Belyaev, S.~Moretti, G.~M. Pruna, and C.~H. Shepherd-Themistocleous.
\newblock {Phenomenology of the minimal B-L extension of the Standard Model}.
\newblock {\em PoS}, EPS-HEP2009:242, 2009.

\bibitem{Basso:2009zz}
L.~Basso, A.~Belyaev, S.~Moretti, G.~M. Pruna, and C.~Shepherd-Themistocleous.
\newblock {Z-prime discovery potential at the LHC in the minimal B-L model}.
\newblock In {\em {6th Les Houches Workshop on Physics at TeV Colliders}}, pages 117--128, 6 2009.

\bibitem{Basso:2009hf}
Lorenzo Basso, Alexander Belyaev, Stefano Moretti, and Giovanni~Marco Pruna.
\newblock {Probing the Z-prime sector of the minimal B-L model at future Linear Colliders in the e+ e- ---\ensuremath{>} mu+ mu- process}.
\newblock {\em JHEP}, 10:006, 2009.

\bibitem{FileviezPerez:2009hdc}
Pavel Fileviez~Perez, Tao Han, and Tong Li.
\newblock {Testability of Type I Seesaw at the CERN LHC: Revealing the Existence of the B-L Symmetry}.
\newblock {\em Phys. Rev. D}, 80:073015, 2009.

\bibitem{Basso:2010dq}
Lorenzo Basso, Alexander Belyaev, Stefano Moretti, Giovanni~Marco Pruna, and Claire~H. Shepherd-Themistocleous.
\newblock {$Z'_{B-L}$ discovery potential at the LHC for $\sqrt{s}=7$ TeV}.
\newblock {\em PoS}, ICHEP2010:381, 2010.

\bibitem{Basso:2010si}
Lorenzo Basso, Stefano Moretti, and Giovanni~Marco Pruna.
\newblock {The Higgs sector of the minimal $B-L$ model at future Linear Colliders}.
\newblock {\em Eur. Phys. J. C}, 71:1724, 2011.

\bibitem{Basso:2010jm}
Lorenzo Basso, Stefano Moretti, and Giovanni~Marco Pruna.
\newblock {A Renormalisation Group Equation Study of the Scalar Sector of the Minimal B-L Extension of the Standard Model}.
\newblock {\em Phys. Rev. D}, 82:055018, 2010.

\bibitem{Basso:2010tv}
L.~Basso, A.~Belyaev, S.~Moretti, and G.~M. Pruna.
\newblock {The Z' boson of the minimal B-L model at future Linear Colliders in e+e- --\ensuremath{>} mu+mu-}.
\newblock {\em Nuovo Cim. C}, 33N2:171--174, 2010.

\bibitem{Basso:2010jt}
L.~Basso, A.~Belyaev, S.~Moretti, and G.~M. Pruna.
\newblock {Tree Level Unitarity Bounds for the Minimal B-L Model}.
\newblock {\em Phys. Rev. D}, 81:095018, 2010.

\bibitem{Li:2010rb}
Tong Li and Wei Chao.
\newblock {Neutrino Masses, Dark Matter and B-L Symmetry at the LHC}.
\newblock {\em Nucl. Phys. B}, 843:396--412, 2011.

\bibitem{Abdallah:2015uba}
W.~Abdallah, J.~Fiaschi, S.~Khalil, and S.~Moretti.
\newblock {Mono-jet, -photon and -Z signals of a supersymmetric (B \ensuremath{-} L) model at the Large Hadron Collider}.
\newblock {\em JHEP}, 02:157, 2016.

\bibitem{Abdallah:2015hma}
W.~Abdallah, J.~Fiaschi, S.~Khalil, and S.~Moretti.
\newblock {Z'-induced invisible right-handed sneutrino decays at the LHC}.
\newblock {\em Phys. Rev. D}, 92:055029, 2015.

\bibitem{Accomando:2016rpc}
Elena Accomando, Luigi Delle~Rose, Stefano Moretti, Emmanuel Olaiya, and Claire~H. Shepherd-Themistocleous.
\newblock {Novel SM-like Higgs decay into displaced heavy neutrino pairs in U(1)' models}.
\newblock {\em JHEP}, 04:081, 2017.

\bibitem{Accomando:2017qcs}
Elena Accomando, Luigi Delle~Rose, Stefano Moretti, Emmanuel Olaiya, and Claire~H. Shepherd-Themistocleous.
\newblock {Extra Higgs boson and Z$^{\prime}$ as portals to signatures of heavy neutrinos at the LHC}.
\newblock {\em JHEP}, 02:109, 2018.

\bibitem{Gutierrez-Rodriguez:2024nny}
A.~Guti\'errez-Rodr\'\i{}guez, J.~M. Mart\'\i{}nez-Mart\'\i{}nez, and M.~A. Hern\'andez-Ru\'\i{}z.
\newblock {The new $Z'$ boson of the B-L model as a portal to signatures of heavy-neutrinos at the future muon collider}.
\newblock {\em Eur. Phys. J. Plus}, 139(6):470, 2024.

\bibitem{Martinez-Martinez:2023qjt}
J.~M. Mart\'\i{}nez-Mart\'\i{}nez, A.~Guti\'errez-Rodr\'\i{}guez, and M.~A. Hern\'andez-Ru\'\i{}z.
\newblock {The Future Muon Collider for the Study of the $Z'$ Boson of the $U(1)_{B-L}$ Model}.
\newblock {\em Int. J. Theor. Phys.}, 62(10):220, 2023.

\bibitem{Iso:2009nw}
Satoshi Iso, Nobuchika Okada, and Yuta Orikasa.
\newblock {The minimal B-L model naturally realized at TeV scale}.
\newblock {\em Phys. Rev. D}, 80:115007, 2009.

\bibitem{Dev:2021qjj}
P.~S.~Bhupal Dev, Bhaskar Dutta, Kevin~J. Kelly, Rabindra~N. Mohapatra, and Yongchao Zhang.
\newblock {Light, long-lived B \ensuremath{-} L gauge and Higgs bosons at the DUNE near detector}.
\newblock {\em JHEP}, 07:166, 2021.

\bibitem{Khalil:2006yi}
Shaaban Khalil.
\newblock {Low scale $B$ - L extension of the Standard Model at the LHC}.
\newblock {\em J. Phys. G}, 35:055001, 2008.

\bibitem{Khalil:2010iu}
Shaaban Khalil.
\newblock {TeV-scale gauged B-L symmetry with inverse seesaw mechanism}.
\newblock {\em Phys. Rev. D}, 82:077702, 2010.

\bibitem{El-Zant:2013nta}
Amr El-Zant, Shaaban Khalil, and Arunansu Sil.
\newblock {Warm dark matter in a $B-L$ inverse seesaw scenario}.
\newblock {\em Phys. Rev. D}, 91(3):035030, 2015.

\bibitem{Basso:2012gz}
Lorenzo Basso, Ben O'Leary, Werner Porod, and Florian Staub.
\newblock {Dark matter scenarios in the minimal SUSY B-L model}.
\newblock {\em JHEP}, 09:054, 2012.

\bibitem{T2K:2019bcf}
K.~Abe et~al.
\newblock {Constraint on the matter\textendash{}antimatter symmetry-violating phase in neutrino oscillations}.
\newblock {\em Nature}, 580(7803):339--344, 2020.
\newblock [Erratum: Nature 583, E16 (2020)].

\bibitem{NOvA:2021nfi}
M.~A. Acero et~al.
\newblock {Improved measurement of neutrino oscillation parameters by the NOvA experiment}.
\newblock {\em Phys. Rev. D}, 106(3):032004, 2022.

\bibitem{Gronau:2011cf}
Michael Gronau and Jonathan~L. Rosner.
\newblock {Triple product asymmetries in $K$, $D_{(s)}$ and $B_{(s)}$ decays}.
\newblock {\em Phys. Rev. D}, 84:096013, 2011.

\bibitem{Bevan:2014nva}
A.~J. Bevan.
\newblock {C, P, and CP asymmetry observables based on triple product asymmetries}.
\newblock 8 2014.

\bibitem{Moretti:2019ulc}
Stefano Moretti and Shaaban Khalil.
\newblock {\em {Supersymmetry Beyond Minimality: From Theory to Experiment}}.
\newblock CRC Press, 2019.

\bibitem{Elsayed:2011de}
A.~Elsayed, S.~Khalil, and S.~Moretti.
\newblock {Higgs Mass Corrections in the SUSY B-L Model with Inverse Seesaw}.
\newblock {\em Phys. Lett. B}, 715:208--213, 2012.

\bibitem{Elsayed:2012ec}
A.~Elsayed, S.~Khalil, S.~Moretti, and A.~Moursy.
\newblock {Right-handed sneutrino-antisneutrino oscillations in a TeV scale Supersymmetric B-L model}.
\newblock {\em Phys. Rev. D}, 87(5):053010, 2013.

\bibitem{Abdallah:2014fra}
W.~Abdallah, S.~Khalil, and S.~Moretti.
\newblock {Double Higgs peak in the minimal SUSY B-L model}.
\newblock {\em Phys. Rev. D}, 91(1):014001, 2015.

\bibitem{Hammad:2015eca}
A.~Hammad, S.~Khalil, and S.~Moretti.
\newblock {Higgs boson decays into $\gamma \gamma$ and Z$\gamma$ in the MSSM and the B-L supersymmetric SM}.
\newblock {\em Phys. Rev. D}, 92(9):095008, 2015.

\bibitem{Hammad:2016trm}
A.~Hammad, S.~Khalil, and S.~Moretti.
\newblock {LHC signals of a B-L supersymmetric standard model CP -even Higgs boson}.
\newblock {\em Phys. Rev. D}, 93(11):115035, 2016.

\bibitem{Abdallah:2016vcn}
W.~Abdallah, A.~Hammad, S.~Khalil, and S.~Moretti.
\newblock {Search for Mono-Higgs Signals at the LHC in the B-L Supersymmetric Standard Model}.
\newblock {\em Phys. Rev. D}, 95(5):055019, 2017.

\bibitem{DelleRose:2017hvy}
Luigi Delle~Rose, Shaaban Khalil, Simon J.~D. King, Carlo Marzo, Stefano Moretti, and Cem~S. Un.
\newblock {Supersymmetric Gauged B-L Model of Dark Matter and Fine Tuning}.
\newblock {\em PoS}, EPS-HEP2017:067, 2017.

\bibitem{DelleRose:2017ukx}
Luigi Delle~Rose, Shaaban Khalil, Simon J.~D. King, Carlo Marzo, Stefano Moretti, and Cem~S. Un.
\newblock {Naturalness and dark matter in the supersymmetric B-L extension of the standard model}.
\newblock {\em Phys. Rev. D}, 96(5):055004, 2017.

\bibitem{Ahmed:2020lua}
Waqas Ahmed, Shabbar Raza, Qaisar Shafi, Cem~Salih Un, and Bin Zhu.
\newblock {Sparticle spectroscopy and dark matter in a $U(1)_{B-L}$ extension of MSSM}.
\newblock {\em JHEP}, 01:161, 2021.

\bibitem{Biswas:2018yus}
Anirban Biswas, Sandhya Choubey, and Sarif Khan.
\newblock {Inverse seesaw and dark matter in a gauged B-L extension with flavour symmetry}.
\newblock {\em JHEP}, 08:062, 2018.

\bibitem{Abdallah:2019svm}
Waleed Abdallah, Sandhya Choubey, and Sarif Khan.
\newblock {FIMP dark matter candidate(s) in a B-L model with inverse seesaw mechanism}.
\newblock {\em JHEP}, 06:095, 2019.

\bibitem{Kanemura:2014rpa}
Shinya Kanemura, Toshinori Matsui, and Hiroaki Sugiyama.
\newblock {Neutrino mass and dark matter from gauged $U(1)_{B-L}$ breaking}.
\newblock {\em Phys. Rev. D}, 90:013001, 2014.

\bibitem{Blanchet:2009bu}
Steve Blanchet, Z.~Chacko, Solomon~S. Granor, and Rabindra~N. Mohapatra.
\newblock {Probing Resonant Leptogenesis at the LHC}.
\newblock {\em Phys. Rev. D}, 82:076008, 2010.

\bibitem{Racker:2008hp}
Juan Racker and E.~Roulet.
\newblock {Leptogenesis, Z-prime bosons, and the reheating temperature of the Universe}.
\newblock {\em JHEP}, 03:065, 2009.

\bibitem{Langacker:2007ur}
Paul Langacker, Gil Paz, Lian-Tao Wang, and Itay Yavin.
\newblock {A T-odd observable sensitive to CP violating phases in squark decay}.
\newblock {\em JHEP}, 07:055, 2007.

\bibitem{Staub:2015kfa}
Florian Staub.
\newblock {Exploring new models in all detail with SARAH}.
\newblock {\em Adv. High Energy Phys.}, 2015:840780, 2015.

\bibitem{Porod:2003um}
Werner Porod.
\newblock {SPheno, a program for calculating supersymmetric spectra, SUSY particle decays and SUSY particle production at e+ e- colliders}.
\newblock {\em Comput. Phys. Commun.}, 153:275--315, 2003.

\bibitem{Porod:2011nf}
W.~Porod and F.~Staub.
\newblock {SPheno 3.1: Extensions including flavour, CP-phases and models beyond the MSSM}.
\newblock {\em Comput. Phys. Commun.}, 183:2458--2469, 2012.

\bibitem{Alwall:2011uj}
Johan Alwall, Michel Herquet, Fabio Maltoni, Olivier Mattelaer, and Tim Stelzer.
\newblock {MadGraph 5 : Going Beyond}.
\newblock {\em JHEP}, 06:128, 2011.

\bibitem{Benedikt:2022kan}
M.~Benedikt et~al.
\newblock {Future Circular Hadron Collider FCC-hh: Overview and Status}.
\newblock 3 2022.

\bibitem{Sjostrand:2014zea}
Torbj\"orn Sj\"ostrand, Stefan Ask, Jesper~R. Christiansen, Richard Corke, Nishita Desai, Philip Ilten, Stephen Mrenna, Stefan Prestel, Christine~O. Rasmussen, and Peter~Z. Skands.
\newblock {An introduction to PYTHIA 8.2}.
\newblock {\em Comput. Phys. Commun.}, 191:159--177, 2015.

\bibitem{deFavereau:2013fsa}
J.~de~Favereau, C.~Delaere, P.~Demin, A.~Giammanco, V.~Lema\^\i{}tre, A.~Mertens, and M.~Selvaggi.
\newblock {DELPHES 3, A modular framework for fast simulation of a generic collider experiment}.
\newblock {\em JHEP}, 02:057, 2014.

\bibitem{fastjet}
Matteo Cacciari, Gavin~P. Salam, and Gregory Soyez.
\newblock {FastJet User Manual}.
\newblock {\em Eur. Phys. J. C}, 72:1896, 2012.

\bibitem{Cacciari:2008gp}
Matteo Cacciari, Gavin~P. Salam, and Gregory Soyez.
\newblock {The anti-$k_t$ jet clustering algorithm}.
\newblock {\em JHEP}, 04:063, 2008.

\bibitem{smlbook}
Andreas Lindholm, Niklas Wahlstr\"om, Fredrik Lindsten, and Thomas~B. Sch\"on.
\newblock {\em Machine Learning - A First Course for Engineers and Scientists}.
\newblock Cambridge University Press, 2022.

\bibitem{TMVA:2007ngy}
Andreas Hocker et~al.
\newblock {TMVA - Toolkit for Multivariate Data Analysis}.
\newblock 3 2007.

\bibitem{FREUND1997119}
Yoav Freund and Robert~E Schapire.
\newblock A decision-theoretic generalization of on-line learning and an application to boosting.
\newblock {\em Journal of Computer and System Sciences}, 55(1):119--139, 1997.

\bibitem{book}
Steven Eppinger and Tyson Browning.
\newblock {\em Design Structure Matrix Methods and Applications}.
\newblock 01 2012.

\bibitem{Cowan:2010js}
Glen Cowan, Kyle Cranmer, Eilam Gross, and Ofer Vitells.
\newblock {Asymptotic formulae for likelihood-based tests of new physics}.
\newblock {\em Eur. Phys. J. C}, 71:1554, 2011.
\newblock [Erratum: Eur.Phys.J.C 73, 2501 (2013)].

\bibitem{Cowan:2017}
Glen Cowan.
\newblock {Estimating Statistical Significance}.
\newblock \url{https://www.pp.rhul.ac.uk/~cowan/stat/cowan_berkeley_4jan17.pdf}, 2017.
\newblock [Online].

\bibitem{Zimmermann:2016puu}
Frank Zimmermann, Michael Benedikt, Xavier Buffat, and Daniel Schulte.
\newblock {Luminosity Targets for FCC-hh}.
\newblock In {\em {7th International Particle Accelerator Conference}}, page TUPMW037, 2016.

\bibitem{Das:2021esm}
Arindam Das, P.~S.~Bhupal Dev, Yutaka Hosotani, and Sanjoy Mandal.
\newblock {Probing the minimal U(1)X model at future electron-positron colliders via fermion pair-production channels}.
\newblock {\em Phys. Rev. D}, 105(11):115030, 2022.

\bibitem{Yin:2021rlr}
Xinyue Yin, Honglei Li, Yi~Jin, Zhilong Han, and Zongyang Lu.
\newblock {Investigating the Z' gauge boson at future lepton colliders *}.
\newblock {\em Chin. Phys. C}, 46(5):053106, 2022.

\end{thebibliography}

\end{document}